Review Article

# How to design research-aligned diversity, equity, and inclusion interventions in physics

Angela Johnson [*]

*Department of Educational Studies, St. Mary's College of Maryland, 47645 College Drive, St. Mary's City, Maryland 20686, USA*

Physicists have called for "a culture and climate that not only is harassment-free but includes people taking active steps to support those who are minoritized … learning spaces that … provide positive experiences where everyone is supported and valued" [A. V. Knaub *et al.*, Changing culture and climate to prevent sexual harassment in the physics educational setting, Phys. Teach. **58**, 352 (2020)]. In this paper, I provide guidance on how to design DEI interventions to support this goal. I address common critiques of DEI interventions and the research base supporting the use of interventions, and discuss physics-specific considerations around DEI interventions [especially concerns about (i) dealing with skeptical participants, and (ii) inadvertently making conditions worse for minoritized participants]. I lay out possible outcomes of DEI interventions, including: increasing participants' awareness of bias; teaching participants to overcome bias; increasing motivation to reduce bias; increasing physics identity; providing recognition to physics students; and increasing the sense of belonging among physicists from minoritized groups. I explore how to justify an intervention to participants and the benefits of making the intervention voluntary or required, and provide possible intervention activities, matched with intended outcomes, as well as a list of physics-specific materials for teaching about stereotype threat, microaggressions, intersectionality, Whiteness, and more. I provide guidance on evaluating interventions, including tools that others have used to measure various outcomes, appropriate quantitative and qualitative approaches, and considerations of rigor. The paper is written in sections; each section begins with a summary and is written so that readers can jump to the section of most interest.

## I. WHAT TO EXPECT IN THIS PAPER

In this article, I lay out evidence for the need for DEI interventions in physics, then examine critiques of DEI interventions and present evidence that these interventions do work, albeit modestly. I show evidence that DEI interventions in physics are typically more well-received than the people responsible for the interventions expected. I then lay out information around a series of questions that a reader might have when designing or choosing a DEI intervention for their setting, including: Who will the participants be? What are the intended outcomes of the intervention? What form will it take? How will it be justified to participants? Will it be voluntary or required? What activities will it consist of? How can it be evaluated to see if it met its outcomes? It is not necessary to read the entire article, or to read it start-to-finish, to make use of it. Each section begins with a sentence or two summarizing the points of interest to readers, and many end with my perspective, based on my experience and my best interpretation of the research. Readers can jump to the sections of most interest.

The intended audience for this paper is individuals or small groups who want to undertake diversity interventions in physics settings but may not be able to institute organization-wide initiatives. This might be a professor who wants to incorporate a focus on DEI into their coursework, a department chair who would like to provide a workshop on DEI to their faculty, or a lab group exploring ways to make their climate more inclusive. Thus, the paper is focused on interventions that take fewer resources to implement: coursework for students; one-off workshops directed at students, faculty, or other members of physics settings. The principles for designing and evaluating interventions like these also apply to larger, systemic interventions, but the examples I use in this article are all drawn from workshops and components of high school and undergraduate physics coursework.

What you will not find in this paper: How to persuade resistant colleagues about the value of DEI interventions. For this, I refer readers to Shaun Harper's suggestions about how to use data to defend diversity, equity, and inclusion [1]. The paper is also not focused on organizational-level reforms; for guidance on this, I recommend the TEAM-UP task force report "The time is now" [2]. The article does not address individual or group coaching, communities of practice, learning communities, or other longitudinal interventions aimed at professionals (though I address longitudinal interventions for students that take place over the course of a semester). For guidance on what these models could look like in physics, I direct readers to the Faculty Online Learning Community model as one example [3].

[*]Contact author: acjohnson@smcm.edu

Finally, this article is not grounded in a particular theory of change. Instead, I wrote it around the participatory action research model, in which people identify a problem in their setting, research how that problem might be addressed, carry out an intervention, and then gather data to see whether the intervention worked [4,5]. I set out to write this paper to answer the questions that physicists often ask me about how to create more inclusive classrooms, departments, and organizations; a classic example of how action research begins. This paper is an attempt to gather together the materials for action research: typical problems around DEI in physics settings; research about the causes of those problems, as well as potential solutions, in order to design reasonable interventions; and information about how to assess whether the intervention worked. I gathered up information on critiques of DEI interventions, experimental approaches to evaluating their effectiveness, and peer-reviewed accounts of DEI interventions in physics. My goal was to get a sense of what we know about DEI interventions in general, in STEM settings, and in physics settings, in order to help physicists design or choose interventions that have a higher likelihood of meeting their goals.

I come to this work after decades of first teaching physics (at the high school level and then as part of the Minority Arts and Sciences Program at the University of Colorado, Boulder) and then researching the experiences of women in science settings. Although much of my research focuses on Black, Latina, and American Indian women, I am a queer White woman. I am not a physicist myself, despite teaching it for fifteen years; I'm a professor of teacher education. My early research interest focused on the dynamics of exclusion of women of color in science settings, particularly at predominantly White institutions [6–8]. My more recent focus has been on university physics settings in particular, and on characteristics and practices at institutions where women (again, especially women of color) report feeling supported and identified with physics [9–12]. Because of these experiences, I have been asked many times to offer workshops on DEI in physics settings. This led me to dive into the research on what actually makes a difference in increasing people's commitment to behaving without bias, an exploration that led to this paper. I have been revising it in the midst of public attacks on supporters of inclusion. Although these attacks have claimed the mantle of fairness, it seems to me that true fairness lies in leveling the physics playing field. Physics is hard enough; no aspiring physicists should face extra challenges just because of the bodies they inhabit. But for now, this article is my effort to stay the course. I hope it will serve your effort to do the same.

## II. WHY SHOULD WE DO DEI INTERVENTIONS IN PHYSICS?

The warped demographics of physics have been amply demonstrated [13–16]. Physicists and those who study physics stress the need to create physics settings centered on diversity, equity, and inclusion, both to recruit physics students and professionals from minoritized groups and to create a better experience for people already in those settings. Physicists have called for "a culture and climate that not only is harassment-free but includes people taking active steps to support those who are minoritized … learning spaces that … provide positive experiences where everyone is supported and valued" [17] and for departments to take "a tough and vocal stance regarding their refusal to tolerate a hostile work environment toward women" [18]. Well done, these efforts can work: a study of women of color and LGBTQ+ students majoring in physics at Texas State found that "widely shared discussions around equity, department structures supporting collaboration, and a culture of teamwork helped them achieve success together" [19]. A Latina at California State University Long Beach described her physics major thus: "Like, yeah, it's hard, but I don't care. It brings me satisfaction actually working through the problems … gave me time to like, go back and learn a lot of stuff that I couldn't pick up if I had not done physics" [20]. A Latina physics major in 2020 requested this from physics professors: "1. Tell your students diversity is important. 2. Give your students resources to understand the underrepresentation and inclusion problem. 3. Emphasize forming a community and valuing others. 4. Remind students that they can learn from each other. 5. Reach out to underrepresented students directly" [21].

So what would this look like, done well? What does the research tell us about effective diversity, equity, and inclusion efforts? In a 2014 article about the need for DEI programming to be grounded in research, the authors argue that "[w]ithout a scientific approach to diversity interventions, we are likely perpetuating the existing system, which fails to uphold meritocratic values by allowing persistent biases to influence evaluation, advancement, and mentoring of scientists," and that "[w]e may also inadvertently continue to fund ineffective interventions that—at best—superficially address diversity goals without producing measurable results, or—at worst—intensify biases" [22]. The intent of this article is to provide that scientific approach; to acquaint physicists with current research on what actually works to create more equitable, inclusive, and perhaps more diverse physics environments.

## III. DEFINITIONS AND TERMINOLOGY

*In this section, I lay out my definition of terms that occur frequently in the rest of this paper.*

Physics education research, equity research, and research into DEI interventions are all replete with multiple terms that refer to the same or very similar concepts. In this paper, I have tried to use only one term for each concept. I chose terms that worked well for me, as a writer, in meeting my goals for this paper at this moment. I am not arguing that these terms are the best choices for the concepts I assigned them to, only trying to clarify in order to make the task of absorbing this information easier by reducing the associated jargon. Here are the terms:

- Diversity, equity, inclusion (DEI): Three values that characterize a setting where people from a wide variety of backgrounds and experiences feel they belong and can thrive.
- DEI intervention: Any intervention provided to members of a setting with the goal of advancing diversity, equity, and inclusion in the setting. The DEI interventions I have profiled in this article take four common forms: light-touch interventions, workshops, stand-alone units in high school and college physics classes, and semester-long themes in physics classes. See Table II for more information about the types of DEI intervention I discuss in this paper, and the Appendix for exemplars of each type:
  - Light touch DEI interventions: very brief activities intended to have long-lasting effects.
  - DEI workshops: one-off events, typically 1 to 3 h in length.
  - DEI units: stand-alone units embedded in high school and undergraduate physics courses.
  - DEI semester-long themes: activities threaded throughout the course of a high school or university physics course.
- Diversity training (DT): This is not a term I have used in my own writing in this paper; however, it is commonly used in the literature I have quoted, usually to refer to a workshop designed to promote inclusion and antibias actions in a setting.
- Physics setting: Any setting where the members regard themselves as physicists, aspiring physicists, or "physics people."
- Physics group member: A person affiliated with a physics setting.
- Minoritized group member/member of a minoritized group: A person who self-identifies or is identified by others as belonging to a group that receives less opportunities for economic security and self-sufficiency. I prefer this to "under-represented" and "minority," because "minoritized" indicates that other people are doing the minoritizing (whether through individual action or through support for institutional inequities). "Marginalized" functions in a similar way, but I don't like the way it seems to push some people onto the margins of a setting, rather than indicating that they are full members of the setting but may have unjustly received fewer opportunities due to identifying or being identified with particular groups. Most of the literature on DEI interventions focuses on women (often approaching gender from a binary perspective), on Black and Latino underrepresentation, and to a lesser extent on LGBTQ+ underrepresented. Dimensions that are seldom addressed but could have a profound impact on a person's experiences in a physics setting include: disability, Indigenous status, social class, country of origin (and, in the United States, related legal status), first-generation college student status, religious affiliation, language proficiency, whether a person is a parent, and others.
- Underrepresented group member or member of an underrepresented group: I have used this term when I want to explicitly remind the reader about the skewed demographics of most physics settings.
- Evaluation: A process of systematically gathering and analyzing data in order to determine whether an intervention met its intended goals.
- Implicit bias: A bias we hold at the implicit (unconscious) level; these biases may be in alignment with or in direct contradiction to our explicit beliefs. Note that implicit bias, unconscious bias, and implicit attitudes are three closely related terms; I leave it to others to distinguish among them. For the purposes of this paper, I am referring to all three concepts under the umbrella term implicit bias.

## IV. CRITIQUES OF DEI INTERVENTIONS

*DEI interventions, particularly short-term workshops, have been criticized for being ineffective and, worse, for having unintended negative consequences* [23–27]. *The validity of this criticism is difficult to assess because most DEI interventions are not studied rigorously* [28]. *I take this to mean that it is easy to do DEI interventions in a shortsighted way that causes harm, but, with research as a guide, it is possible to design and implement effective interventions.*

Let us pause to look into the difficulties of a "scientific approach to diversity interventions" [22]. These critiques typically focus on short-term DEI workshops (abbreviated as DT in some quotes below); however, some of the critiques apply to larger DEI interventions as well. Critics argue that DEI workshops are "likely the most expensive, and least effective, diversity program around," because (i) "short-term educational interventions in general do not change people"; (ii) "antibias training activates stereotypes"; (iii) "training inspires unrealistic confidence in anti-discrimination programs, making employees complacent about their own biases"; (iv) "training leaves whites feeling left out"; and (v) "people react negatively to efforts to control them" [23].

An intervention like this can backfire in several ways. It may incorrectly signal that an institution is more inclusive than it is, or "signal that underrepresented groups need help to succeed and are thus less competent than their advantaged counterparts" [24]. Short-term DEI interventions may even make things worse: "Communicating that the solution to bias is simple and can be addressed by just offering a training course can decrease empathy for victims of bias. Additionally, referring to bias as ubiquitous may, ironically, produce the perception that bias is acceptable (for instance, 'It must be OK everyone has it') or that bias-reduction efforts are futile (for instance, 'You can't succeed because bias is too widespread to root out')" [25]. Researchers have found that DEI interventions have led to "increased belief in race essentialism, or the notion that racial group differences are valid, biologically based, and immutable" [26] and that a focus on implicit bias can result in the belief that people who engage in discriminatory behaviors are less accountable because their bias is unconscious [27].

Evaluating these critiques is complicated due to the absence of evidence; often, there is a mismatch between the stated goals of a DEI intervention and the outcomes that are measured. Researchers reviewed 250 reports on DEI interventions and found that common goals included retention of minoritized employees, improving perceptions of workplace climate, decreasing the frequency of workplace discrimination, improving climate for members of historically minoritized groups, and mitigating disparities in educational outcomes [28]. However, outcome measures were typically completion of the intervention, whether participants liked the intervention, whether participants' awareness of DEI issues increased, and whether participants' self-efficacy around DEI increased.

This mismatch between goals and measured outcomes makes it difficult to know whether DEI interventions are working. The researchers argue that "we should be ethically bound to demonstrate that DT programs are effective and, just as important, do no harm" and that "greater attention should be paid to the types of outcomes that will provide evidence that the DT offered is actually effective" [28].

My take: Critiques of DEI interventions are grounded in important insights; rather than being deterred from carrying out a DEI intervention on the basis of these critiques, the critiques can help shape a more effective intervention, one which is designed to avoid harm and which is rigorously evaluated for its effectiveness.

## V. MODEST EXPERIMENTAL SUPPORT FOR DEI INTERVENTIONS

*Experimental studies indicate that DEI interventions lead to modest but statistically significant decreases in implicit and explicit bias and positive changes in behavior* [29,30]. *Impacts were stronger when the interventions were one aspect of a larger institutional commitment to DEI* [31]. *The evidence suggests that DEI interventions can increase the likelihood that a physics department will commit to other DEI efforts* [32].

These critiques of DEI interventions make two recent meta-analyses particularly useful. These are analyses of DEI interventions that were tested using experimental procedures, and in which the stated goals of the interventions were measured explicitly [29,30]. These analyses give reason to think that DEI interventions can be worth doing, particularly when accompanied by (or serving as an impetus for) wider institutional transformation efforts. In both meta-analyses, researchers identified studies measuring whether diversity interventions reduced implicit and explicit biases. Forscher and colleagues also looked at whether behavior changed after the experimental intervention. In both meta-analyses, researchers compared various activities intended to reduce bias and found that some activities are more effective than others; I summarize these findings later, in the section called "What activities will you use?" Each group of researchers found that overall, attempts to change implicit and explicit attitudes produced significant but small changes.

The effect of an intervention is measured by effect size, which is the difference between the mean of the intervention group and the control group on the variable under study, compared to the variation in the control group [33]. For context, a large effect size is 0.8; medium is 0.5; and small is 0.2 [34]. Forscher and colleagues found effect sizes that averaged 0.3. Paluck and colleagues found larger effect sizes, averaging 0.37. However, both groups of researchers found evidence that these values may be artificially inflated due to publication bias. Paluck and colleagues report that the reduction in bias among the studies in the top quintile of sample sizes (and thus the least susceptible to producing outcomes only due to chance) drops to just 0.187. Both research groups also found little to no connection between changes in implicit and explicit biases: a diversity intervention aimed at reducing participants' implicit bias will not decrease their explicit bias. Forscher and colleagues also found that reductions in implicit bias were not associated with changes in behavior [29].

These modest impacts, however, still result in some level of prejudice reduction, and other researchers remind us that institution-wide changes must be accompanied by changes in individuals and result from individual actions. "Despite increased focus on structural and systemic forms of discrimination, with accompanying mantra, 'individual solutions can't solve structural problems,' we contend that they can, and, further, that structural solutions that deemphasize individual psychology do so to their detriment" [35]. A controlled study of 92 STEM departments (46 of which received a research-based DEI intervention) found that "[i]nstitutional transformation requires changes at multiple levels, yet it is the individuals within an institution who drive change. Consistent with these tenets of institutional change, we found that an intervention that helped individual faculty members change their gender bias habits led to positive changes in perception of department climate: increased perceptions of fit, valuing of research, and comfort in raising personal and professional conflicts" [31]. Bringing a few more individuals on board through DEI interventions—especially if they are individuals in a position to influence department practices and culture—can leverage wider change.

Furthermore, modest effect sizes may, under the right circumstances, lead to bigger impacts. In reflecting on the small effect sizes but detectable positive results of anti-bias workshops, Carnes *et al.* point out that

> "tiny effect sizes can be amplified through: (1) repetition (in our case, a faculty member from a minoritized group might interact with many colleagues practicing bias-reducing strategies), (2) the environment in which the intervention occurs (in our case, departments were frequently investing in other pro-diversity initiatives which could amplify the small effect of [their intervention]), (3) downstream consequences with greater impact than the initial effect (in our case, more respectful division meetings might enhance faculty retention), and (4) the scaling up of tiny effect sizes when large numbers of people are engaging in the new behavior (in our case, if many members of a division role model bias-reducing strategies, it might positively impact the training of future physicians)" [32].

These points are based on a review article on how to make sense of effect sizes in context, in which the authors lay out conditions under which small effect sizes can have outsize real-world implications through the four dynamics cited above [36]. In my view, this means that the relatively small effect sizes of the activities I lay out below nonetheless have the potential to create meaningful differences in a physics setting, and thus it is even more important to do them well the first time.

DEI interventions are more effective (and small effect sizes are more likely to grow into measurable change) when they are a part of an overall commitment to DEI; "[t]he key to improving the effects of training is to make it part of a wider program of change" [23], whereas "[i]f the organization lacks accountability, efforts to promote inclusion will likely be in vain" [37]. In a review of 250 articles evaluating DEI interventions in organizational, human services, and educational settings, the authors found that "individual DT [diversity training] was more impactful on surrogate outcomes when delivered alongside larger workplace diversity initiatives" [28]. Using DEI interventions in addition to, rather than as a substitute for, wider efforts can "help ensure that the billions spent each year yield meaningful change" [25].

My take: DEI interventions can result in at least some individuals becoming more committed to diversity, equity, and inclusion, and thus can increase the likelihood that a physics department will commit to other DEI efforts.

## VI. PHYSICS-SPECIFIC CONCERNS AROUND DEI INTERVENTIONS

Many physicists have expressed concern that participants will resist DEI interventions. However, those who have carried out such interventions have experienced minimal pushback; typically, 75%–100% of participants have welcomed the experience [38–41]. Another concern is putting participants from minoritized groups in the spotlight. Again, this seems unfounded; women of color, in particular, have expressed appreciation for well-designed DEI interventions [38,42,43].

It has been my experience that physicists are concerned about their own awkwardness and possible pushback from students if they institute DEI interventions. Moses Rifkin, a high school physics teacher who has been instrumental in developing the Underrepresentation Curriculum Project, raised the same concern in an email to me: "I think the fear of retribution and the fear of stumbling/making a mistake are the two biggest fears among facilitators … [however,] anecdotally, we haven't heard very much at all about people getting pushback" [41]. Research reports on DEI interventions in physics contexts confirm Rifkin's experience. In all the studies that I could locate that reported students' views after experiencing DEI interventions in physics settings, the majority of participants did not object to it, and substantial numbers fervently welcomed it. In one study, 75%–80% of student responses were positive, 15%–20% were neutral, and 0% (one year) and 5% (the next year) were negative. "The few negative reactions that we have received have primarily included the idea that these topics are important, but they just do not belong in a physics class" [38]. Daane and Rifkin reported that 96% of high school students found that a DEI unit was worthwhile, and 90% of college students said that a similar unit was a positive experience [39]. In another study, Daane quotes a student saying, "My thoughts regarding this whole situation have gradually changed. I went in thinking that racial diversity in the physics classroom was not important at all. I now think that diversity can be a great and empowering thing not just to the minority but to the whole classroom" [40].

Another concern involves the possibility that DEI interventions will inadvertently put a spotlight on participants from minoritized groups, who may already be feeling isolated in physics settings. Anecdotal evidence indicates that with forethought, this, too, can be avoided. Dalton and Hudgings report that after their entire class participated in a DEI unit, "two female Students of Color who were interviewed described feeling less pressure to 'prove' their experiences with bias or inequity, to themselves or their peers, because these issues were being addressed more generally, in a less directly personal context for them" [42]. A woman who participated in a DEI unit as part of her high school physics class reported that "After these classes, I would say that there was a change. Big changes happen over time, so of course, nothing became perfect, but people seemed to be more aware of their actions, and there was less tension, because we had all talked about what we were experiencing" [43]. Baylor *et al.* reported that students from minoritized groups felt more like they belonged in physics as a result of a DEI unit in their physics class [38]. A woman in their study said, "I have never been confident that I belong to the physics community. [The reflection essays] taught me to keep challenging myself about the stereotypical views of physicists: it does not have to be white/male/genius/people from academic families/smartest kid in class/etc., all it takes is really a passion for physics." A woman of color went further: "This class opened my eyes. It may have been one of the most influential classes I have attended to this day."

My take: I personally find these results comforting. They address my two biggest worries every time I stand outside the door of a physics classroom, waiting to go in and begin a DEI intervention: That I will face outright resistance from some students and that I will inadvertently make things worse for the minoritized students in the classroom. Given the small-but-non-zero positive outcomes from carefully designed experimental DEI interventions, coupled with the general acceptance that physicists and physics educators have found and the strong positive reactions from minoritized students, I hope that what follows—detailed guidelines from the research base on how to design DEI interventions for physics contexts—will be of use to other colleagues who share my fears and commitments.

In the remainder of this article, I have laid out research-based principles for guiding the development of DEI interventions in physics contexts. The principles are derived from the two meta-analyses of experimental activities included in DEI interventions and the review of DEI evaluations I described above [28–30], as well as two research-based models for DEI interventions [22,35] and 15 published reports on 11 different DEI interventions carried out in physics settings (see the Appendix). I have broken the guidance up into questions to make it as accessible as possible; I spend the remainder of the paper answering them:

1. Who will your participants be? Faculty or other long-term members of physics organizations? Students? How will you take into account skeptical participants? What about participants from minoritized groups?
2. Why are you doing this intervention? What outcomes do you hope for?
3. What form will your intervention take? A light touch (e.g., a short activity during one class session)? A workshop? A unit that takes several days within a physics course? A theme that runs throughout a physics course?
4. How will you justify the intervention to participants?
5. Will the intervention be voluntary or required?
6. What activities will you use?
7. How will you know whether your intervention worked?

## VII. WHO WILL YOUR PARTICIPANTS BE?

The first question to consider is whether your intervention is aimed at faculty (or other long-term members of physics organizations) or students.

### A. Faculty

*DEI interventions for faculty typically take the form of a workshop, seldom grounded in research* [28]. *I recommend using this article or other resources to ensure that your interventions are grounded in research about what activities actually increase diversity, equity, and inclusion in physics settings.*

According to the review of DEI interventions by Devine and Ash [28], interventions in organizational settings like university departments typically have goals like "the recruitment and retention of employees from minoritized backgrounds as well as increased group cohesion, creativity, and equity within a given workplace." The intervention often consists of a DEI workshop: a lecture provided by an outside consultant, followed by group activities like reviewing cases of discrimination (indeed, I myself have delivered this sort of workshop). Devine and Ash report that "the selection of particular DT strategies appears to be most often motivated by personal preference or intuition about what trainers believe would be effective rather than by a specific theoretical approach or empirical evidence" and that there is not a strong evidence base about the effectiveness of this kind of DEI intervention, chiefly because when these interventions are evaluated, the evaluations are not typically aligned with the goals of the intervention.

My take: It would be worthwhile for departments to either design research-based workshops using the information in this article or to ask outside consultants to show the research basis for their workshops.

### B. Students

*There is some research on creating assignments and units that actually increase support for DEI goals in physics classes; I strongly urge readers to consult this research.*

In the same review, Devine and Ash [28] report that DT directed towards students typically has goals including "to improve school climate for members of historically marginalized groups and to mitigate disparities in educational outcomes." Forms can range from one-off activities or assignments to units to semester-long courses; in the Appendix, I include physics examples of all of these. Again, student-focused interventions are seldom evaluated rigorously or against their own intended outcomes.

### C. Skeptics and resistant participants

*To address potential skepticism or resistance from participants, research indicates that DEI initiatives should include both carefully-designed DEI interventions and also formal policies around promoting diversity (and remediating discrimination). Interventions should*:

- ask participants at the beginning of the intervention to reflect on what they appreciate about themselves, the institution they belong to, and their own moral commitments to justice [44],
- emphasize "shared responsibility for addressing diversity challenges" rather than blame [45], and
- consider creating external motivations for behavioral change rather than trying to change participants' fundamental beliefs [35].

There is a helpful body of research about how to design DEI interventions that will mitigate disruptions from skeptics, resistant participants, and what one group of researchers calls "low empathy individuals" [46]. The Underrepresentation Curriculum website has an excellent resource, Working With Skeptics, which addresses not only how to work effectively with skeptical participants but also how to respect one's own skepticism and that of parents and administrators (particularly relevant to DEI interventions carried out with high school participants). The advice: "listen openly to the concerns expressed by skeptics …. Behind the concern and emotion are likely legitimate fears and worries that deserve acknowledgement" [47].

Shuman *et al.* [48] have shown that people who resist DEI interventions do so because they feel threatened; they may be experiencing a status threat (that gains of members of minoritized groups will come at their expense), merit threat (that if they accept that bias and discrimination have held back members of minoritized groups, it will diminish their own success), and moral threat (that if they acknowledge the privilege of members of their group, they are morally diminished because they are linked to an unfair system). In response to these threats, people may engage in denying the existence of injustice, defending the current system (a classic example is the memo from a former Google employee that argued that women are underrepresented among coders not because of discrimination but because of inherent differences between women and men [49]), or distancing themselves—arguing that while discrimination may exist, they have never perpetrated nor benefited from it.

Shuman and colleagues suggest that status threat can be mitigated by making the "business case" for DEI. (I suggest this tentatively, given the findings that the business case can backfire with both majority group members and members of minoritized groups [44]. For more information, see the section below called "How will you justify the intervention to participants?") They suggest that merit threat can be mitigated by self-affirmations; "instead of beginning a meeting about the need for diversity training by providing statistics about the severity of the problem, consider first engaging people in an exercise allowing them to reflect and affirm themselves, or highlighting positives about the organization and its employees that provide this sense of affirmation." (I have done this myself by beginning DEI workshops by asking participants what they value about physics and the physics community). Finally, they suggest that moral threat can be mitigated by "reframing DEI interventions as a way for people to express their moral ideals and thus repair their moral standing," an approach that aligns well with suggestions in the section that follows, "Increase participants' awareness of their own biases/the impact of bias on others."

One group of researchers suggests that "low empathy individuals" actually need DEI interventions more than others: "while high empathy individuals are attuned to the needs of diverse populations and thus internally motivated to respond without prejudice toward them, low empathy individuals may require diversity training to promote motivation and ultimately reduce prejudice" [46]. Gill and Olson [35] suggest that the motivation might need to be external, and describe what this could look like: "Organizations might foster external motivation by putting in place reward structures for achieving goals related to diversity—and punishments, such as the potential to be fired, when people actively violate egalitarian expectations by prejudicial speaking or active discriminating." This approach, they argue, will be most effective with people who identify strongly with the group, establishing such expectations.

My take: In my experience, both in physics and teacher education environments, the emphasis seems to be on changing people's internal beliefs, but I see great utility in simply trying to reduce overt discrimination.

### D. Participants from minoritized groups

*Members of minoritized groups are, ironically, at risk from participating in DEI interventions* [28]. *They may be harmed by being put in an unwelcome spotlight and treated as members of a group rather than full humans* [21,26] *or learning that other members of physics settings hold stereotypes about people like them* [21,50–52] *or regard the interventions as a waste of time* [40]. *They may also be harmed if other members of their setting feel threatened by the interventions* [48]. *These threats can be minimized by basing interventions in evidence* [38]; *focusing on shared responsibility for creating inclusive environments rather than individual blame* [31,45]; *allowing people to remain* *quiet or participate privately in interventions* [38]; *and careful preparation and structuring of the intervention* [53]. *These approaches also reduce resistance to DEI interventions.*

Any DT that is intended to improve the experiences of physicists from minoritized groups needs to take those individuals' needs into account in its design, in order to "ensure no further harm is done to students from marginalized backgrounds in the process of the discussion" [54]. The people at the Underrepresentation Curriculum website are direct on this point: "When in doubt, we encourage instructors to consider prioritizing the needs of the marginalized over the needs of the majority" [47].

I found clear statements of the importance of this commitment and our responsibility to do it well (e.g., "instructors need support and preparation in implementing equity and social justice in classrooms supporting students of various demographic backgrounds" and should "seek guidance from others with this expertise and/or resources that have strategies for doing this" [54]). I also found research on the dangers of poorly designed DEI initiatives: "in considering the utility of intergroup contact as a tool for increased inclusion on college campuses, further attention should be allocated to the experiences of contact for people of color. Some research suggests that the positive effects of intergroup contact may not extend to members of historically marginalized groups" [28].

It seems to me that DEI initiatives could cause harm to participants from minoritized groups by (and this will by no means be a complete list) (i) putting them in a spotlight, where they are asked to speak as representatives of their group; (ii) causing majority group members to feel blamed for inequity ("moral threat"), that doubt has been cast on their successes ("merit threat"), or that their positions are threatened ("status threat"), and thus to resent minoritized group members [48]; (iii) to feel essentialized, reduced to just avatars of the groups rather than full humans [21,26]; (iv) to get their stereotype threat fears confirmed by learning that members of their community really do have lower expectations of them based on their group membership [21,50–52]; (v) to have to choose between correcting inaccurate information (as an example, one research participant told me that a professor used the term Chicano to indicate anyone with Latino/Hispanic origins, when the term more commonly indicates Latino/Hispanic folks who hold a particular political orientation) presented by poorly informed DEI leaders (thus drawing attention to themselves as members of minoritized groups) or tolerating the inaccurate information (and possibly suffering consequences from it); (vi) to have to listen to classmates devalue DEI ("It really bothered me hearing my classmates complain about this [topic] in physics. Most of my classmates say they want a change in how we perceive minorities, yet they are complaining that they even have to talk about it. How is a person going to make a change or even give ideas [for] change if he/she is not willing to face this uncomfortable topic?" [40]). This list is not comprehensive; I am certain there are other ways that members of minoritized groups can be harmed by participating in DEI interventions that do not center on their needs.

Fortunately, these harms can be mitigated by ensuring that DEI interventions are rooted in evidence. For example, interventions can rely on information from AIP, AAPT, APS, and other professional organizations [38]. Interventions can focus on community members' shared responsibility for creating a healthy climate rather than on assigning individual blame, and present DEI work as a shared moral effort to advance justice [31,45]. Interventions can also allow for private ways of participating, or for non-participation; for example, DEI interventions that are threaded through a semester can allow participants to submit their reflections in writing (as assignments) rather than requiring group discussion in order to participate [38]. Members of minoritized groups can be told in advance about what to expect in upcoming interventions, so that they may prepare accordingly [40,55]. The DEI interventions can focus on participants' shared identity as members of the physics community rather than their separate identities as majority or minoritized group members [56].

More ideas can be found in an excellent set of suggestions on how to discuss sexual harassment in physics settings (written after the publication of an article on the prevalence of sexual harassment in physics settings [57]). The authors provided a list of questions to consider before carrying out an intervention, including "Can you envision an outcome of women and gender minorities participating that would be significant enough that they might choose to open themselves to this potential for harm?" [53] They also included a list of suggestions for carrying out the intervention, including "In advance of a discussion, make sure participants have a good understanding of what they should expect in the discussion, so they can mentally prepare. At the beginning of the discussion, explicitly set ground rules for conversation" and "Consider alternative formats for discussions, such as having groups where men can discuss [the article on harassment] and what they will do about it without asking or expecting women and gender minorities to be present" [53]. The National Equity Project also has an excellent collection of tools and guides for organizations and individuals seeking to create more equitable environments [58]. Their Developing Community Agreements guidelines are particularly relevant to designing interventions that honor participants from minoritized groups.

My take: Follow this advice from Devine and Ash: "[R]esearch on DT [diversity training], as well as its practice, only infrequently attends to the perspectives and experiences of individuals who are at risk for experiencing discrimination. Given this focus, we suggest that historically marginalized individuals should be consulted during the planning process if DT is to be effective in meeting its goals. Input and involvement from members of historically marginalized groups should be actively sought in determining whether and how to deliver DT within their settings" [28]. Happily, many of the activities that can make DEI interventions more beneficial to people from minoritized backgrounds can also address skeptics; rather than having to choose between the needs of the minoritized or the resistant, it is sometimes possible to address their needs together.

## VIII. WHY ARE YOU DOING THIS INTERVENTION? WHAT OUTCOMES DO YOU HOPE FOR?

Your intended outcomes will guide the rest of your design. Gill and Olson recommend that you "choose a target for intervention that makes sense: implicit bias; awareness of impacts of bias; internal motivation to reduce bias; external motivation to reduce bias; opportunity to override implicit attitudes in order to take bias-reducing actions" [35]. In this section, I present research on possible outcomes; see Table I for a summary of the outcomes, as well as reasonable activities to bring about these outcomes and instruments to assess whether the outcomes were met.

### A. Reduce implicit bias

*While some interventions have been shown to reduce participants' implicit biases (as measured on the Implicit Association Test, for example), these changes do not lead to changes in either explicit bias or behavior* [28–30]. *Thus, I do not recommend reduction of implicit bias as an outcome goal unless it is accompanied by other goals; perhaps not even then, as it is not clear that implicit bias is, in fact, a measure of participants' animus toward other groups. There is reason to think it might be a measure instead of the cultural norms of their environment* [29]. *Furthermore, focusing on implicit bias can have unintended consequences, including reducing participants' views on the seriousness of discrimination* [27] *and leading participants to think that bias is normal and, therefore, acceptable* [35].

Reducing implicit bias is an area where there is a strong research base, presumably both because it is intuitively appealing (if our implicit biases shape our behavior, then it stands to reason that changing those biases will change our behaviors) and because it is easy to measure using implicit attitude tests. Unfortunately, despite the intuitive appeal, reductions in implicit bias have not been shown to be related to changes in behavior or reductions in explicit bias. In their meta-analysis of experimental attempts to reduce bias, Forscher and colleagues "found little evidence that changes in implicit measures translated into changes in explicit measures and behavior, and we observed limitations in the evidence base for implicit malleability and change" [59]. Paluck and colleagues found that "across studies, there appears to be no correlation ($r = 0.02$) between estimated effects on implicit prejudice and estimated effects on explicit prejudice [28,30].

Forscher and colleagues propose an explanation as to why reducing implicit biases doesn't seem to affect behavior or explicit attitudes [59]. They liken implicit biases to a scar. Rather than thinking of implicit biases as drivers of biased behavior, they suggest that "automatically retrieved associations reflect the residual 'scar' of concepts that are frequently paired together within the social environment and do not have much causal force on their own." If this is the case, then "implicit measures could be used to predict the prevalence of certain judgments or behaviors within a population," but "efforts to change behavior by changing implicit measures would be misguided." Effort would be better spent on trying "to rid the social environment of the features that cause biases on behavioral and cognitive outcomes" or helping people develop "strategies to resist the environment's biasing influence." With this model, we might expect people in physics settings to manifest high levels of implicit biases towards, for example, Black women and Latinas (who together make up around 2% of the people who graduate with bachelor's degrees in physics), but their implicit biases might just reflect what we already know about the demographics of physics, not anything about the individuals who hold those biases.

In addition to the research indicating that changing implicit biases does not seem to change behavior or explicit bias, focusing on implicit bias can have unintended consequences. It can reduce the seriousness with which discriminatory behavior is viewed by others.

> [P]erpetrators of discrimination are held less accountable and often seen as less worthy of punishment when their behavior is attributed to implicit rather than to explicit bias. Moreover, at least under some circumstances, people express less concern about, and are less likely to support efforts to combat, implicit compared with explicit bias [27].

Worse, teaching about implicit bias can lead people to regard bias as normal, and thus feel more free to express prejudice [35]. Furthermore, a focus on implicit bias may distract from "other pathways to discrimination, and, by extension, how DT might produce change in other ways" [35].

My take: do not use implicit bias reduction as your central intended outcome of your DEI intervention; maybe do not use it at all.

### B. Increase participants' awareness of their own biases/the impact of bias on others

*Increasing participants' awareness of their own bias (in a context that avoids assigning blame for inequity)* [45,48]

*and the impact of bias on others* [35], *when coupled with opportunities to develop strategies to overcome bias* [28], *has been shown to have results that persist for several years. These outcome measures, in combination, can result in long-lasting changes in some participants.*

Rather than focusing just on implicit bias, Carnes and colleagues recommend that DEI interventions focus on increasing participants' awareness of their own bias in order to promote changes in behavior [31]. Gill and Olson present evidence that implicit bias may not even be fully unconscious; that "[w]hen implored to be honest, people can report on explicit measures their attitudes revealed by implicit measures" [35]. Thus, they advise that "[i]nstead of ill-informed discussions of 'unconscious bias,' DT practitioners should focus on how automatic forms of bias can lead to unintentional discrimination, particularly when there is either ambiguous or abundant information on which to base a judgment."

If you choose this outcome, research suggests that there are effective and ineffective activities to include in your intervention. As Dobbin and Kalev remind us, "people react negatively to efforts to control them" [23]. An initial research report from the Network Contagion Research Institute and the Rutgers Social Perception Lab suggests that when DEI interventions "focus heavily on victimization and systemic oppression," this can lead participants to overshoot and read racist motivations into neutral scenarios [60]. Finally, participants who are skeptical of DEI interventions are unlikely to respond well to being told to examine their own morality for signs of failure. I share this view; it seems to me that when the goal is to increase respect and inclusion for minoritized community members, it does not make sense to approach other members of the same community disrespectfully, and it is fundamentally disrespectful to demand that people search themselves and find themselves individually wanting.

This can be solved, however, by representing bias not as an individual flaw but as a human experience, and one that we, as humans, have the power to overcome. For example, Shuman, Knowles and Goldenberg recommend that this be mitigated through "reframing DEI initiatives as a way for people to express their moral ideals and thus repair their moral standing," citing research showing that "when DEI initiatives are framed as a way to express universal ideals (fairness, equality, and so on) rather than as an obligation that majority-group members must live up to, this increases support for DEI programs" [48]. In a workshop that significantly increased professional biologists' willingness to take action on DEI issues, "the workshop emphasized shared responsibility for addressing diversity challenges and avoided assigning blame for diversity issues … the blame for diversity challenges does not lie with members of specific groups" [45].

Increasing DEI intervention participants' awareness of their own biases is often coupled with increasing their knowledge about how bias impacts others. Awareness of the impacts of bias can be a particularly powerful outcome because research indicates that "most White individuals are surprisingly uninformed about the extent to which racial inequality persists" [35]. This may be particularly the case in physics settings; in an article outlining how "settler colonial white supremacy, capitalism, and United States imperialism all contributed to create the field of physics we have today," the authors point out that "physicists commonly are unaware of this history, and many do not value social justice and equity work" [54].

Care must be taken in how to educate participants about the impact of bias, however. Hideg and Wilson found that when the focus is on historical impacts, this can actually decrease support for present-day equity policies, by "fostering the belief that inequality no longer exists" [61]. Devine and Ash found that "diversity courses that target students' knowledge and awareness without attending to mechanisms of behavioral change are likely not sufficient to create lasting changes in the form of reduced expressions of bias, increased intergroup inclusion, and improved feelings of belonging for marginalized students" [28].

My take: knowledge of the history of bias (without making it clear that its impacts are ongoing) can backfire, and knowledge about the present-day impacts of bias is not a sufficient outcome; if you include information about the history and impacts of bias in your intervention, be sure you couple it with with training about how to improve the situation, the topic covered in the next section.

### C. Increase participants' skills in overcoming bias

*Effective DEI interventions first help participants become aware of their own bias and of the impact of bias and then help participants develop or practice strategies to overcome bias* [31,59,62]. *The evidence for this outcome is presented below; information on how to help participants acquire bias-reducing strategies is presented in the section entitled "Teach strategies to overcome bias," under "What activities will you use?"*

Developing strategies to overcome bias is another useful outcome. In the section on activities, I discuss how to do this in an intervention. There is strong experimental support for increasing participants' awareness of their own biases, their awareness of the impact of bias on others, and their skill with strategies to overcome bias. "The most effective training is anti-bias training that is designed to increase awareness of bias and its lasting impact, plant seeds that inspire sustained learning, and teach skills that enable attendees to manage their biases and change their behavior" [25]. These three outcomes fit well conceptually for participants and, in combination, have been shown to change participants' behaviors for as long as two years.

In a preregistered experimental study by Carnes and colleagues, faculty from a random selection of departments at one university attended a DEI workshop, while faculty from nonselected departments were not invited [31]. The workshop was designed to "first increase faculty's awareness of gender bias in academic medicine, science, and engineering, and then to promote motivation, self-efficacy, and positive outcome expectations for habitually acting in ways consistent with gender equity." For experimental departments in which at least 25% of the faculty attended the workshop, faculty were significantly more likely to report that they fit into their departments better, their colleagues valued their research more, and they were more comfortable bringing up family and personal responsibilities in scheduling department events than members of departments in the control group. Effect sizes for these differences ranged from 0.11 to 0.32 [31]. In the two years following the intervention, about 32% of the new faculty hired by the control departments were women vs 47% of the new faculty hired in the experimental departments [62]. In a subsequent study, this intervention was scaled up to 19 departments of medicine in 16 different states, with similar results: small but statistically significant improvements in participants' awareness of their own bias and in their familiarity with and use of strategies to reduce bias. Furthermore, COVID-19 created a natural experiment that allowed the researchers to confirm that they got similar results when the workshop was delivered in person and online [32].

Further support comes from an experimental study by Forscher and colleagues [59]. In an intervention designed to break participants' "prejudice habit," Forscher and colleagues offered participants in the intervention group a workshop that

- increased participants' knowledge through "giving participants feedback about their own level of implicit bias, as measured by the Implicit Association Test,"
- taught them "how implicit bias can lead to unintentional but consequential discriminatory behavior, leading to negative consequences for racial minorities," and
- provided them with "evidence-based, cognitive strategies that, if practiced, can lead to bias reduction."

They argue that these activities worked because, unlike implicit biases, changes in knowledge may result in longer-lasting behavioral changes, because knowledge "is flexible enough to be influenced by new information and central enough to the self-concept to support the continuation of intervention-initiated changes." The researchers found that two weeks following the workshop, participants in the intervention group were more likely to notice bias, label bias as wrong, and have interracial interactions with others. Two years after the workshop, participants were more likely to disagree publicly with an essay arguing that stereotypes were harmless (48% of those in the control condition posted a public comment disagreeing with an essay, compared with 79% of the intervention group).

My take: Increasing participants' skill at overcoming bias is an excellent outcome for DEI interventions.

### D. Increase opportunities to practice strategies to overcome bias

*It is not enough to develop strategies to overcome bias; people need the opportunity to use those strategies* [35]. *Especially for DEI interventions aimed at those in leadership positions (faculty members; student leaders, for example, of SPS), this might be a useful outcome.*

Gill and Olson point out that DEI interventions focused on increasing opportunity to act without bias "have received little attention; the only examples we have seen are the occasional recommendations to 'slow everything down' in hiring processes to combat the effects of bias" [35]. They argue that factors that interfere with this opportunity include stress, time pressure, poor sleep, burnout, and substance abuse. While leaders in physics settings cannot necessarily address substance abuse, I can imagine a department working together in the process of a DEI intervention to create decision-making structures and departmental interaction norms that are protected from stress and time pressure, and department policies and cultural norms that help avoid burnout and all-nighters.

### E. Increase internal or external motivation to reduce bias

*While DEI interventions typically focus on increasing participants' internal motivations, there are people who will not be internally motivated; external motivations in the form of antibias cultural norms and policies can change the behavior of these individuals without relying on changing their underlying beliefs* [35].

DEI interventions often focus on increasing internal motivation to behave without bias (and in the section below on interventions, I present a range of techniques that have been shown to do this with some degree of effectiveness). Gill and Olson remind us, however, that some people will not be internally motivated. "For people who are indifferent to inequality and discrimination or who are motivated to maintain the *status quo*, interventions aimed at external motivation may at least reduce overt discrimination" [35]. If you choose increasing external motivation as a way to reduce bias in your setting, they recommend that "[w]hen conducted in group settings, DT is a fitting context to increase external motivation through making nonprejudicial norms salient" (particularly, I would add, DEI interventions directed at faculty members or department leadership). However, they caution against being too heavy-handed; "organizational interventions that threaten autonomy often backfire, undermining motivation and increasing implicit and explicit expressions of prejudice" [35]. In "The Time is Now," the TEAM-UP task force has some specific suggestions for what this might look like in a university physics department: "If a department has a major [DEI] initiative, it is likely that the chair has used incentives and/or rewards to foster that initiative. These may include course releases compensating for the time faculty put into advising students or launching the initiative, seeking seed funding, and advocating with the dean or provost for such funding; they may also include departmental awards or nominations for external awards and prizes" [2].

My take: Ideally, our DEI interventions would increase participants' intrinsic motivation to reduce the bias in their behaviors, but barring that, a reduction in biased behaviors due to extrinsic motivation is a lot better than no change in participants' behaviors.

### F. Increase physics identity/intent to pursue a physics major or career

*Increasing students' physics identity has been pursued extensively; research suggests that interest in physics careers among women can be increased through the use of counternarratives discussing women's underrepresentation in physics and including women physicists in the curriculum. Such approaches have been shown to significantly increase the intentions of women and students from minoritized groups to pursue physics* [63,64]. *Research into the role of recognition in fostering physics identity underscores its importance; however, less is known at this point about how to systematically pursue this outcome* [6,8,65–67].

Increasing students' physics identity (their sense that they are a "physics person") has been shown to predict students' achievement and retention in physics, and their career interests in physics [2]. Hazari and colleagues studied five classroom experiences to see their relationship to girls' interest in careers in physical science: "(i) having a single-sex physics class, (ii) having female scientist guest speakers, (iii) having a female physics teacher, (iv) discussing the work of female scientists, and (v) discussing the underrepresentation of women" [63]. Based on results from a survey of a national sample of 7505 students in college English classes at 34 colleges and universities across the United States, the researchers established that discussion of underrepresentation was highly significantly correlated with interest in a career in physical science. Discussing the work of female scientists trended very strongly toward significance ($p = 0.05$). As is typical of attempts to address bias and inequity, the effect sizes were small: for discussions of underrepresentation, the effect size was 0.27, and for discussions of the work of women scientists, the effect size was 0.18.

In an experimental investigation of the impact of counternarratives about who does physics (using the STEP-UP lesson Careers in Physics) and why they do it (using the STEP-UP lesson Women in Physics), researchers found that compared with a control group, both women and students from minoritized groups showed significantly higher gains in intent to pursue physics as a result of participating in the lessons. Effect sizes were 0.29 (for female students) and 0.30 for students from minoritized groups [64].

In my own work, the science identity of women of color majoring in the sciences was deeply tied to whether they were recognized by others as a science person, including whether the recognition they received was positive or negative [6,8]. Singh and colleagues have documented the power of recognition on women in physics contexts. In a qualitative study of 38 women taking physics, Li and Singh found that 25 reported negative recognition (including "feeling belittled for questions or efforts," "Feeling negatively recognized about their abilities and potential" and feeling marginalized when men dominated in class or men and women were treated differently in class) [66]. They also found forms of positive recognition, including having their ability recognized, being encouraged to pursue their goals, and having struggles in physics framed as normal. However, only four students in their sample reported positive recognition. In a quantitative study of 827 students taking physics, Cwik and Singh found that women's perceptions of whether their physics instructors and teaching assistants positively recognized them as "a physics person" was correlated with their final course grades in physics [65]. In a quantitative study with 1045 physics students, Li and Singh found that students' perceived recognition from physics instructors predicted their physics identity [67]. Given the relationship between perceived positive recognition in physics and physics identity, training for physics instructors and teaching assistants in how to provide positive recognition to physics students (and how to avoid negative recognition) seems prudent. I have not found research on interventions that support physics instructors in learning to offer positive recognition; if anyone has done this work, please tell me about it.

My take: First, it is indeed possible to increase girls' and women's interest in physics, but several of the common-sense ways people hope to do this are not well-supported by research; if this is an intended outcome of your DEI intervention, choose discussion of underrepresentation as one of your activities. Note that the discussion of underrepresentation is very similar to raising participants' awareness of the impacts of bias. Second, recognition of students' ability and seeing students as "physics people" can have profound impacts on students' performance in physics and their physics identity, and is another useful outcome to consider.

### G. Increase the sense of belonging/decrease the sense of isolation of physics students who are members of minoritized groups

*Although research on how to increase a sense of belonging is still tentative, I recommend this as an*

*outcome, because 1) members of minoritized groups in physics frequently report that they feel like they don't belong* [21,50,57,68–71] *or are harassed and excluded* [72] *(and research supports that this feeling goes along with lower persistence and performance* [67,68,73–75]), *and 2) it is possible to create physics settings where members of minoritized groups report they feel they belong* [9,19] *or experience less exclusion* [76]

This outcome is particularly salient to physics communities, as there is evidence that (i) a sense of belonging is associated with success in STEM [77,78] and physics [2,67,68,73–75] and (ii) members of minoritized groups often expcerience a lower sense of belonging than other students in STEM [78,79] and physics [2,21,50,57,68,69]. There is also evidence that women of color in physics sometimes experience intense isolation [70,71] and that LGBTQ+ physicists (particularly trans physicists) experience exclusionary behaviors that are so intense they consider leaving [72].

As usual, this outcome is more likely to be met when it is part of wider inclusion efforts. A light touch DEI intervention designed to increase students' sense of belonging only increased their persistence when it was accompanied by larger opportunities for belonging (for example, students and faculty from diverse groups; clubs, events, and coursework that support minoritized students' group identities; pedagogy organized around growth mindsets; and campus settings where positive intra- and inter-group interactions occur [80]). This is supported by the finding that women feel a weaker sense of belonging in STEM when they believe that STEM faculty show gender bias [81]. A well-designed DEI intervention to increase women's sense of belonging is unlikely to work if the women who participate in the intervention believe their professors are biased against them. Within physics, this can be done by emphasizing "the nature of community and belonging within the scientific practices of the classroom" [68].

Although I was not able to identify research in which DEI interventions in physics measurably increased the sense of belonging of minoritized students, I found good descriptions of what it feels like to belong in a physics class, as well as research correlating overall positive climates in physics settings with fewer exclusionary experiences for minoritized physicists [76]. This suggests that this outcome is possible, even if it requires more than just a one-off DEI workshop to bring it about. From a study carried out at UT San Antonio: "… your fellow classmates, like, they don't–they're not there to watch you fail. They want you to succeed. They want to help you. Everyone's pulling each other along, because we're all suffering together." "It's very friendly and it's a community and it's a very tight knit department. […] I have a really good support system that has made me a lot happier in school than I was in high school … we all have common interests and we're really good friends and we have this community built around supporting each other to the best of our abilities." "… I just love the sense of a community. I don't know. It's- 'community.' That's a good word. … It's a really good sense of community, and I've really found my place in it [19]." From my own research: "physics is what I've always been interested in. It doesn't feel like I'm out of place. It's the subject I'm interested in. So I don't really think about it" (from a Black woman in her third year); "I mostly love it. It's been a really great experience majoring in physics here. It's been really really hard but I love it!" (from a mixed-race woman in her senior year) [9].

My take: Since it is possible for minoritized students to feel a sense of belonging in their physics departments, this is a superb goal for departments to pursue. I see no downside to trying to create a sense of community among physics students.

TABLE I. Reasonable activities to achieve particular outcomes.

| Intended outcome | Reasonable activities |
| --- | --- |
| Reduce implicit bias | Create real or imaginary contact<br>Teach about inequity<br>Investigate who belongs and what it means to belong in physics |
| Increase participants' awareness of their own or others' bias | Teach strategies to overcome bias<br>Facilitate discussion<br>Invoke values and beliefs |
| Increase participants' awareness of the impacts of bias on others | Create real or imaginary contact<br>Teach about inequity<br>Facilitate discussion |
| Increase participants' skills with overcoming bias | Teach strategies to overcome bias<br>Create opportunities for action |
| Increase opportunities to practice strategies to overcome bias | Facilitate discussion<br>Create opportunities for action |
| Increase motivation to reduce bias and willingness to take action | Create real or imaginary contact<br>Teach about inequity<br>Facilitate discussion<br>Invoke values and beliefs |
| Increase physics identity and intent to pursue a physics major or career | Create real or imaginary contact<br>Investigate who belongs and what it means to belong in physics<br>Facilitate discussion<br>Invoke values and beliefs |
| Increase belonging and decrease isolation of physicists who are members of minoritized groups | Teach about inequity<br>Investigate who belongs and what it means to belong in physics<br>Facilitate discussion<br>Invoke values and beliefs<br>Create opportunities for action |

TABLE II. Categories of DEI interventions.

| Type | Definition and experimental support | Examples |
|---|---|---|
| Light touch interventions | "Treatments that are easy to implement, brief (under 10 minutes), inexpensive, and thought to have lasting effects" [30].<br>Effect size: 0.16–0.35, but note that "[t]he prejudice reduction literature is largely silent on whether light touch interventions durably reduce prejudice" [30].<br>My own advice: Be very cautious about replicability. | Pro-diversity posters [35]<br>Reflecting on information from more advanced students to ease the transition to college [80] |
| Workshops | One- to three-hour events, typically consisting of one or two sessions. Most workshops are not well evaluated, [28] but a few have been shown experimentally to have results that persist for up to 2 years [22,31,62]. | Typically focus on the skewed demographics of physics, implicit bias, information about the impacts of inequity, and sometimes strategies to combat inequity and exclusion |
| Units within high school or university physics courses | One-day to one-week units (one to five class sessions) focused on DEI in physics. Students typically appreciated the units; [38,39,41,55] one unit on careers in physics and women in physics (the STEP-UP curriculum) resulted in significant increases among all kinds of students in their interest in pursuing physics-based careers [64]. | Underrepresentation curriculum<br>STEP-UP curriculum<br>Discussing the 2016 open letter to SCOTUS from professional physicists |
| Semester-long themes in high school or university physics courses | Periodic classwork or assignments on DEI themes threaded through a semester-long physics class. Research on efficacy is scarce, but two studies indicated that students' views on the value of pro-diversity themes increased across the semester [38,42]. | Readings, essays, and written reflections on DEI themes, including information about diverse physicists in lectures, and assigning problems with diverse settings and actors |

In Table I, I summarize the possible outcomes discussed in this section and activities that have been shown to lead to those outcomes. Guidance on how to measure the outcomes in Table I can be found in Sec. XIII, "How will you know whether your DEI intervention worked?" Extensive discussion of reasonable activities can be found in Sec. XII below, "What activities will you use?"

## IX. WHAT FORM WILL YOUR INTERVENTION TAKE?

I have grouped interventions into four categories: light touch interventions (likely aimed at students and incorporated into a class session early in the semester), workshops (which could be aimed at faculty, at students, or at other participants in physics settings), short units in physics classes (aimed at students), and semester-long themes in physics classes.

The form of your intervention will be determined by your audience and the time you have available for it. The interventions aimed at faculty that I have profiled in this paper all take the form of workshops. Interventions for students can take all four forms. Note, however, that the most well-known light touch intervention in physics, in which women's grades in physics courses were higher if they wrote about their values as a class exercise, has never been replicated [82,83] and furthermore "[t]he prejudice reduction literature is largely silent on whether light touch interventions durably reduce prejudice" [30]. My own take is to be very cautious about the replicability of light-touch interventions. Research supporting the other three forms is stronger.

In Table II, I provide definitions of the four categories of DEI interventions that I found among the physics exemplars I identified, with examples. For more examples, see the Appendix.

## X. HOW WILL YOU JUSTIFY THE INTERVENTION TO PARTICIPANTS?

*I (tentatively) suggest you justify your intervention using a fairness case (diversity is the right thing to do) rather than an instrumental case (diversity leads to better science).*

In physics settings, DEI interventions are frequently justified through what Georgeac and Rattan call the "business case:" "diversity is valuable for organizational performance" [44]. In physics, "organizational performance" gets translated to "better science." They contrast the business case with the "fairness case" ("justifying diversity as the right thing to do"). While it may seem on the surface like the business case would make DEI interventions more palatable for majority group members, these researchers report that in fact "after being exposed to a business (vs fairness) case for diversity, White Americans report more negative beliefs about inclusion, and exhibit more biased decision making towards Black job applicants."

The business case also backfires, perhaps even more spectacularly, with people who are members of minoritized groups. Through a set of five preregistered experimental studies, Georgeac and Rattan showed that when LGBTQ+ professionals, women seeking STEM jobs, African American students, and African American college graduates were presented with hypothetical cases for the importance of diversity, they showed less interest in working for institutions that used the business case than those that used the fairness case. They also anticipated that they would feel a lower sense of belonging if they joined the business case organizations. Differences were small to moderate, as is typical of the research covered in this review. This study focuses on hypothetical justifications for diversity based on the arguments that researchers culled from Fortune 500 websites, rather than justifications given as part of DT.

My take: This research suggests that we use the business case with caution.

## XI. WILL THE INTERVENTION BE VOLUNTARY OR REQUIRED?

*There are drawbacks both to requiring participation in DEI interventions [23] and to making it voluntary [25]. Two ways to thread the needle: (i) focus the intervention on an issue where there is a strong shared interest in improvement [35] and (ii) target the intervention to socially connected people [28,31].*

The main argument against required interventions is that "people react negatively to efforts to control them" [23]. Because of this, Dobbin and Kalev suggest we should "make training voluntary, or give employees a choice of different types of diversity training" [23]. However, although people may react better to voluntary training, "when training is voluntary, behavioral learning is significantly lower compared with when training is mandatory … perhaps because those who could benefit most from the training avoid attendance" [25]. A meta-analysis of 260 diversity interventions found that whether the intervention was required or voluntary had no effect on outcomes [84].

Gill and Olson thread this needle: "our experience is that while some of the choir finds being preached to be validating, others resent being mandated to attend DT and/or being 'told what they already know'" [35]. Because this question is unsettled, they suggest that "in 'saturated markets,' where training is ubiquitous, and in contexts where diversity-related challenges are known and more specific, a more tailored approach makes sense." They also point out that "acts of discrimination are committed primarily by a particularly biased minority of individuals. In short, bias may be pervasive, but it is distributed unevenly." Thus, they recommend that "DT practitioners should investigate the distribution of bias in a given context and decide from there whether their intervention should address bias or another factor." This might include, for example, DEI interventions focused on a particular issue that there is strong agreement about changing; a mix of voluntary and mandatory workshops; a choice of intervention options; or voluntary workshops coupled with changes in policy or procedures.

Devine and Ash suggest another way to thread this needle: "Rather than making DT voluntary or mandatory, consider targeting socially connected individuals within an organization" [28]. The findings of Carnes and colleagues support this: They found that their workshop "worked" as long as at least 33% of a department attended, and it worked whether or not the department chair attended [31]. Research on the diffusion of innovations supports this. "The adoption of an innovation by individuals in an organization is more likely if key individuals in their social networks are willing to support the innovation" [85], and under ideal conditions, if as few as 25%–35% of a group adopt a new social norm, a tipping point can be reached where the norm will spread to the whole group [85].

My take: Rather than requiring attendance at DEI interventions, design interventions that address a problem that matters to many members of your physics setting and make participation as easy as possible.

## XII. WHAT ACTIVITIES WILL YOU USE?

*Choose activities that are matched with the intended outcomes of your intervention.*

In this section, I present some activities that are common in physics DEI initiatives (and relevant experimental research findings when possible), as well as some activities that are not used as commonly but have been shown to be effective in other contexts. I suggest that you choose activities that are well suited to your intended outcomes (see Table I). For example, if your intended outcome is to raise awareness about the impacts of bias on members of minoritized groups, reasonable activities might include real or imagined contact with members of those groups or taking on the perspective of group members. If your goal is to increase participants' motivation to create a more inclusive environment, you might invoke their values around fairness.

### A. Teach about inequity

*There is evidence that teaching about inequity in DEI interventions lowers participants' prejudice levels [28], and also that students from minoritized groups appreciate it [21,40,42,86]. In physics, this might consist of information about the demographics of physics, implicit bias, stereotype threat, beliefs about "natural ability" in physics, benefits of heterogeneous working groups, and the prevalence of sexual harassment in physics.*

Eight of the eleven DEI interventions in physics that I included as exemplars in the Appendix included teaching participants about inequity. Many interventions included information about the demographics of physics settings (often referencing information from AIP, AAPT, APS, and the NSF). Some also included information about imposter syndrome, implicit bias, stereotype threat, beliefs about "natural ability" in physics, structural racism/white privilege, the isolation experienced by members of historically excluded groups in physics settings, affirmative action, meritocracy, and what microaggressions are. Other topics included the benefits of heterogeneous learning and working environments, intersectionality, and how to use an intersectional lens to understand the dynamics of a situation, and the prevalence of sexual harassment in physics settings. For resources that would be helpful in approaching some of these topics, see Table III.

According to Devine and Ash, this kind of DEI intervention is typical in educational settings, and is typically evaluated through pre-post assessment of participants' knowledge and attitudes; of the 51 peer-reviewed research reports of this kind of DEI intervention they located since 2000, only 7 included an experimental design in assessing the effectiveness of the intervention [28]. That said, many of the studies reported a reduction in participants' self-reported levels of prejudice after completing the intervention, or lower levels compared with students who did not participate in the intervention. While this evidence is not particularly strong, at least it points in the right direction, suggesting that teaching about inequity is a reasonable strategy.

Another reason to teach about inequity (what its dimensions are in physics and how it is experienced) is that students from minoritized groups report that they appreciate it. For example, after a unit on equity taught an introductory physics class, one student wrote, "I am a Mexican female. I have experienced discrimination. Therefore, I strongly support discussions inside classrooms about race" [40]. In an article about actions physics instructors can take to change the environment in physics classes, a Latina physics major specifically requests that instructors "give your students resources to understand the underrepresentation and inclusion problem" [21]. "An emphasis by departments and faculty on the idea of physics as something people do, rather than physics as a set of knowledge already created by inaccessible geniuses, can support access for marginalized students to the identity of physicist" [86]. After a semester-long intermediate-level modern physics class that included some of these topics woven throughout the semester, "two female Students of Color who were interviewed described feeling less pressure to 'prove' their experiences with bias or inequity, to themselves or their peers, because these issues were being addressed more generally, in a less directly personal context for them" [42].

My take: Teaching about inequity is a good activity for DEI interventions in physics. It pairs well with a number of powerful DEI intervention outcomes (see Table I), including reducing implicit bias, increasing participants' awareness of bias, increasing their willingness to reduce bias, and increasing the sense of belonging of physicists from minoritized groups.

### B. Create real or imagined contact

Real and imagined contact with people from minoritized groups can decrease biases and increase inclusive behavior [29,30,35]. It must be done carefully so as not to spotlight members of minoritized groups [56]. Imagined contact through storytelling (perhaps biographies or case studies) is a good option in physics [21,99,109,110].

Another useful activity is to create contact between members of well-represented and minoritized groups in physics. This can be done via face-to-face contact, imagined contact, or what Rosa and Mensah call "storytelling" [99]. The mechanism here is that contact with outgroup members can decrease bias by replacing people's negative beliefs and stereotypes with actual positive experiences. In a meta-analysis of 492 experimental studies focused on decreasing bias, researchers found that activities (including real and imagined contact) that indirectly increase participants' positive associations with members of outgroups can decrease implicit and explicit bias and increase inclusive behavior [29].

Face-to-face contact between ingroup and outgroup members has been shown experimentally to decrease bias, with effect sizes in the 0.25–0.28 range [30]. Positive outcomes include improved attitudes towards members of outgroups; increased empathy and commitment to egalitarianism; decreased anxiety about interacting across group lines; and humanizing minoritized group members [35]. This kind of contact fits well with the general importance of strong connections in physics settings; "physics educators should structure classroom practices in ways that maximize students' opportunities to create positive social connections" [69]. In physics settings, however, because of the skewed demographics of physics as well as the danger of spotlighting DEI intervention participants who may already feel minoritized in a physics setting, face-to-face contact should be structured carefully. It might consist of inviting a person from a minoritized group to speak (rather than asking DEI intervention participants to represent their group), or of carefully facilitated discussions in which outgroup members can speak up but are not expected to [56].

Due to the inherent difficulties of in-group and out-group contact in physics settings, another good option is imagined contact. This has also been shown experimentally to lower bias, with effect sizes ranging from 0.12 to 0.37 [30]. "In a prototypical imagined contact intervention, participants are asked to take a few minutes to imagine themselves meeting a member of another group" [30]. In physics, this might consist of reading biographies [109] or learning about the accomplishments [110] of women in physics, or using case studies illustrating the experience of minoritized physics students [56].

TABLE III. Materials for teaching about inequity.

| Topic | Resources |
| --- | --- |
| Demographics of physics settings | • AIP Statistical Research [87].<br>• Physics Education Statistics [88] (From the American Physical Society).<br>• The Underrepresentation Curriculum Project Unit 1, Laying the Foundation [47]. |
| Stereotype threat: Descriptions of what it feels like | • Herrera, S., Mohamed, I. A., & Daane, A. R. (2020). Physics from an underrepresented lens: What I wish others knew. *The Physics Teacher*, 58(5), 294–296 [21].<br>• Doucette, D., & Singh, C. (2020). Why are there so few women in physics? Reflections on the experiences of two women. *The Physics Teacher*, 58(5), 297–300 [50].<br>• Johnson, A., Ong, M., Ko, L. T., Smith, J., & Hodari, A. (2017). Common challenges faced by women of color in physics, and actions faculty can take to minimize those challenges. *The Physics Teacher*, 55(6), 356–360 [70].<br>• Ong, M. (2023). *The Double Bind in Physics Education: Intersectionality, Equity, and Belonging for Women of Color*. Harvard Education Press [71].<br>• The Underrepresentation Curriculum Project Unit 2, Gaining Relevant Knowledge [47]. |
| Microaggressions: What they look like in physics | • Herrera, S., Mohamed, I. A., & Daane, A. R. (2020). Physics from an underrepresented lens: What I wish others knew. *The Physics Teacher*, 58(5), 294–296 [21].<br>• Ong, M. (2005). Body projects of young women of color in physics: Intersections of gender, race, and science. *Social problems*, 52(4), 593–617 [89].<br>• Ong, M., Smith, J. M., & Ko, L. T. (2018). Counterspaces for women of color in STEM higher education: Marginal and central spaces for persistence and success. *Journal of Research in Science Teaching*, 55(2), 206–245 [90].<br>• Johnson, A., & Mulvey, E. (2021). A classroom intervention to promote equity and inclusion. American Physical Society April Meeting [56].<br>• The Underrepresentation Curriculum Project Unit 2, Gaining Relevant Knowledge [47]. |
| "Natural ability" beliefs in physics: impacts | • Leslie, S.-J., Cimpian, A., Meyer, M., & Freeland, E. (2015). Expectations of brilliance underlie gender distributions across academic disciplines. *Science*, 347, 262–265 [91]. |
| Intersectionality in physics: The power of intersectional analyses in physics | • Santana, L. M., & Singh, C. (2024). Importance of accounting for student identities and intersectionality for creating equitable and inclusive physics learning environments. *Phys. Rev. Phys. Educ. Res.*, 20(2), 020142 [92].<br>• Keblbeck, D. K., Piatek-Jimenez, K., & Medina Medina, C. (2024). Undergraduate physics students' experiences: Exploring the impact of underrepresented identities and intersectionality. *Phys. Rev. Phys. Educ. Res.* 20, 020120 [93].<br>• Cochran, G. L., & Bodeva, M. (2020, October 1, 2020). A framework for improving diversity work in physics. Proceedings of the Physics Education Research Conference (PERC) [94].<br>• Quichocho, X. R., Conn, J., Schipull, E. M., & Close, E. W. (2019). Who does physics? Understanding the composition of physicists through the lens of women of color and LGBTQ+ women physicists. *Proceedings of the Physics Education Research Conference (PERC)*, 488–493 [95].<br>• Quichocho, X. R. (2020). Ya basic: Examining the duality of minority-serving conference experiences. *The Physics Teacher*, 58(6), 408–412 [96].<br>• Quichocho, X. R., Schipull, E. M., & Close, E. W. (2020). Understanding physics identity development through the identity performances of Black, Indigenous, and women of color and LGBTQ+ women in physics. *Physics Education Research Proceedings 2020* [86].<br>• Schipull, E. M., Quichocho, X. R., & Close, E. W. (2019). "Success together": Physics departmental practices supporting LGBTQ+ women and women of color. *Proceedings of the Physics Education Research Conference*, 535–540 [19].<br>• Ko, L. T., Kachchaf, R. R., Hodari, A. K., & Ong, M. (2014). Agency of women of color in physics and astronomy: Strategies for persistence and success. *Journal of Women and Minorities in Science and Engineering*, 20(2) [97].<br>• Ong, M., Smith, J. M., & Ko, L. T. (2018). Counterspaces for women of color in STEM higher education: marginal and central spaces for persistence and success. *Journal of Research in Science Teaching*, 55(2), 206–245 [90]. |

TABLE III. *(Continued)*

| Topic | Resources |
| --- | --- |
| | • Bjorkquist, R., Bogdan, A. M., Campbell, N. L., Chessey, M., Cochran, G. L., Cunningham, B., Esquivel, J. N., Gladstone, L., Gosnell, N. M., Guruswamy, S., Hallinen, K. M., Harris, C., Johnson, A., Johnson, J. L., Jones, C., Jorgenson, R. A., McCullough, L., McNeese, M. D., Presley, T. D., … Singh, C. (2019). Women in physics in the United States: Reaching toward equity and inclusion. *AIP Conference Proceedings*, 2109(1), 050040 [51].<br>• Cochran, G. L., Hyater-Adams, S., Alvarado, C., Prescod-Weinstein, C., & Daane, A. R. (2021). Social Justice and Physics Education. In C. C. Ozaki & L. Parson (Eds.), *Teaching and Learning for Social Justice and Equity in Higher Education: Content Areas* (pp. 125–147). Springer International Publishing. [54]<br>• Johnson, A. (2020). An intersectional physics identity framework for studying physics settings. In A. J. Gonsalves & A. Danielsson (Eds.), Physics Education and Gender (pp. 53-80). Springer [9].<br>• The Underrepresentation Curriculum Project Unit 2, Gaining Relevant Knowledge [47]. |
| Sexual harassment in physics settings: | • Aycock, L. M., Hazari, Z., Brewe, E., Clancy, K. B. H., Hodapp, T., & Goertzen, R. M. (2019). Sexual harassment reported by undergraduate female physicists. *Physical Review Physics Education Research*, 15 [57].<br>• Barthelemy, R. S., McCormick, M., & Henderson, C. (2016). Gender discrimination in physics and astronomy: Graduate student experiences of sexism and gender microaggressions. *Physical Review Physics Education Research*, 12(2), 020119 [18]. |
| Isolation: descriptions | • Santana, L. M., & Singh, C. (2023). Effects of male-dominated physics culture on undergraduate women. *Physical Review Physics Education Research*, 19(2), 020134 [98].<br>• Herrera, S., Mohamed, I. A., & Daane, A. R. (2020). Physics from an underrepresented lens: What I wish others knew. *The Physics Teacher*, 58(5), 294–296 [21].<br>• Ong, M. (2005). Body projects of young women of color in physics: Intersections of gender, race, and science. *Social Problems*, 52(4), 593–617 [89].<br>• Ong, M. (2023). The double bind in physics education: Intersectionality, equity, and belonging for women of color. Harvard Education Press [71].<br>• Doucette, D., & Singh, C. (2020). Why are there so few women in physics? Reflections on the experiences of two women. *The Physics Teacher*, 58(5), 297–300 [50].<br>• Quichocho, X. R. (2020). Ya basic: Examining the duality of minority-serving conference experiences. *The Physics Teacher*, 58(6), 408–412 [96].<br>• Ko, L. T., Kachchaf, R. R., Hodari, A. K., & Ong, M. (2014). Agency of women of color in physics and astronomy: Strategies for persistence and success. *Journal of Women and Minorities in Science and Engineering*, 20(2) [97].<br>• Rosa, K., & Mensah, F. M. (2016). Educational pathways of Black women physicists: Stories of experiencing and overcoming obstacles in life. *Physical Review Physics Education Research*, 12(2), 020113 [99]. |
| Affirmative action | • An open letter to SCOTUS from professional physicists [100].<br>• The Underrepresentation Curriculum Project Unit 2, Gaining Relevant Knowledge [47]. |
| Whiteness | • Robertson, A. D., & Hairston, W. T. (2022). Observing whiteness in introductory physics: A case study. *Physical Review Physics Education Research*, 18(1), 010119 [101].<br>• Robertson, A. D., Vélez, V., Hairston, W. T., & Bonilla-Silva, E. (2023). Race-evasive frames in physics and physics education: Results from an interview study. *Physical Review Physics Education Research*, 19(1), 010115 [102].<br>• Dancy, M., & Hodari, A. K. (2023). How well-intentioned white male physicists maintain ignorance of inequity and justify inaction. *International Journal of STEM Education*, 10(1), 1–29 [103].<br>• Dancy, M. (2025). Majority group physics instructors struggle to notice, name, and disrupt racist and sexist student interactions. *Physical Review Physics Education Research*, 21(1), 010142 [104]. |

*(Table continued)*

TABLE III. *(Continued)*

| Topic | Resources |
| --- | --- |
| Structural issues | • Young, R. (2020). Parenting and Physics: How to support physics students who are raising children. *The Physics Teacher*, 58(6), 382–386 [105].<br>• National Academies of Sciences, E., & Medicine. (2024). *Supporting family caregivers in STEMM: A call to action*. The National Academies Press [106].<br>• Kachchaf, R., Ko, L., Hodari, A., & Ong, M. (2015). Career–life balance for women of color: Experiences in science and engineering academia. *Journal of Diversity in Higher Education*, 8(3), 175–191 [107].<br>• Cochran, G. L., Hyater-Adams, S., Alvarado, C., Prescod-Weinstein, C., & Daane, A. R. (2021). Social justice and physics education. In C. C. Ozaki & L. Parson (Eds.), *Teaching and Learning for Social Justice and Equity in Higher Education: Content Areas* (pp. 125–147). Springer International Publishing [54].<br>• The Underrepresentation Curriculum Project Unit 2, Gaining Relevant Knowledge [47]. |
| Implicit bias | • Women in Physics—Step Up [108].<br>• The Underrepresentation Curriculum Project Unit 2, Gaining Relevant Knowledge [47]. |
| Imagined contact: Case studies | • Johnson, A., & Mulvey, E. (2021). A Classroom Intervention to Promote Equity and Inclusion. American Physical Society April Meeting [56].<br>• Careers Quick Reference—Step Up [108]. |

Another activity is the use of storytelling or counter-narrative: "teachers and teacher educators can make use of storytelling to unveil and validate the experiences of students of color in science" [99]. This "gives voice to people from traditionally marginalized groups to share their untold experiences. Counter-storytelling, as the name implies, challenges common social views and ideas" [21]. Paluck and colleagues call this type of activity "entertainment": "interventions that seek to leverage the power of narrative or artistic transportation to overcome natural human tendencies to counterargue messages or resist persuasion attempts. Narrative transportation refers to a psychological phenomenon whereby individuals are carried away by a story, causing them to let down their tendency to question, critique, or counterargue" [30]. The effect size of the twelve examples of entertainment they identified ranged from 0.38 to 0.43.

My take: Creating real or imagined contact can be a powerful tool to evoke participants' motivation to reduce bias in physics. A screening of *Hidden Figures* would be an obvious use of entertainment to provide imagined contact.

### C. Teach strategies to overcome bias

*Strategies to overcome bias (stereotype replacement, counterstereotype imaging, perspective taking, individuation, and seeking out contact with people from minoritized groups* [59,111]*) have been shown experimentally to reduce biased behavior and improve institutional climate for as long as two years* [28,30,31,62]. *Research suggests that perspective taking may be the most powerful of these strategies* [46].

In their meta-analysis, Paluck and colleagues found that activities "which train individuals to use thinking and emotion regulation strategies to fight off their personal prejudices" were successful in reducing bias (effect size for all studies in their sample was 0.35; for those with 78 or more participants and thus greater statistical power, the effect size was 0.22) [30]. Devine and Ash argue for the importance of activities like these: "Incorporating evidence-based prejudice reduction strategies alongside diversity-related course content is likely essential if DT seeks to make meaningful changes in discriminatory behaviors, not just attitudes" [28].

Five particular strategies for overcoming bias have been studied extensively: *stereotype replacement* (noticing when you encounter, even within yourself, a stereotyped assumption, and replacing it with more accurate information), *positive counterstereotype imaging* (noticing stereotypical thoughts and replacing them either with an image of a person, known or imagined, who defies the stereotype), *perspective taking* (imagining what it's like to be a member of a minoritized group; more on this later), *individuation* (taking care to gather information about people in order to guard stereotyped assumptions), and *seek out contact* with outgroup members. These five strategies, in combination, have been shown to reduce bias with effects that persist as long as two years [59,111]. Researchers speculated that use of these five strategies "increases people's sensitivity to bias, particularly when others act with bias, and increases the probability that, when a person encounters bias, he or she will label that bias as wrong" [59].

Carnes and colleagues offered workshops to 46 randomly selected academic medicine, science, and engineering departments at the University of Wisconsin-Madison [31]. The workshops explicitly taught participants to use these five strategies and to write down plans for using them. After the intervention, "faculty in experimental departments felt that they ‘fit in’ better ($p = 0.024$), that their colleagues valued their research and scholarship more ($p = 0.019$), and that they were more comfortable raising personal and family responsibilities in scheduling department obligations ($p = 0.025$) …. Results were consistent across male and female faculty, and workshop attendance by the department chair/head had no impact.” The effect sizes of these differences ranged from 0.11 to 0.32. Positive outcomes again persisted for at least two years; departments that participated in the workshop hired 18% more women faculty than those that were randomly assigned not to participate [62].

So far, I have cited studies that used all five of these strategies together. This leaves us with a question: Do any of the five strategies work well on their own? Do any work better than others? While I did not find research that directly compares these five strategies, a longitudinal study randomly assigned 118 participants to practice one of three strategies: perspective taking, goal setting, and stereotype discrediting (which I understand as a loose proxy for the stereotype replacement, positive stereotype imaging, and individuation strategies) [46]. Eight months after the intervention, participants in the perspective-taking group showed more supportive behaviors toward outgroup members than those in the goal-setting or stereotype-discrediting groups. These researchers found that perspective taking “may be particularly beneficial for individuals who need the training more than others (i.e., individuals who are low in empathy)” [46].

I have incorporated these strategies in a workshop I have offered many times to physics students and faculty at several institutions [56]. In the workshop, I collectively call the three strategies that directly address stereotypes “don’t be a jerk” (language that might be too blunt for your setting, but at my own institution we have a code of conduct that is universally referred to as “DBAJ” so it works for me); I call the perspective-taking strategy “watch good movies and read good books”; and I call seeking out contact “make new friends.” Another example from a physics setting is the guide to bias-reduction strategies around women in physics that is part of the STEP-UP curriculum, Everyday Actions. The tools include a checklist of everyday actions that teachers can use to guard against bias in their interactions with students.

My take: Teaching strategies to overcome bias is the single most important activity to include in a DEI intervention. Why are we doing the intervention except to help participants behave in less-biased ways?

### D. Investigate who belongs and what it means to belong in physics

*This strategy is a good choice for physics settings, given the power of the “lone genius” myth that permeates cultural beliefs about physics* [71]. *Boundaries of who belongs in physics can be expanded through case studies of famous physicists from minoritized groups, through asking all students in a physics class to imagine themselves in physics-based careers* [108] *and by emphasizing what group members have in common (interest in physics!—because it’s interesting!) rather than focusing on differences* [30].

Ong argues that “the solution to underrepresentation in physics is not located in ‘diversity efforts’ or the striving of marginalized groups. Instead, it is found in recognizing the enduring whiteness, maleness, and heteronormativity of physics culture and in dismantling those institutional norms and replacing them with a culture of belonging” [71]. Activities that address belonging were classified as “social categorization” in the meta-analysis by Paluck and colleagues: social categorization activities “encourage participants to rethink group boundaries or to prioritize common identities shared with specific outgroups” [30]. Their thinking behind why this works is that “the simple act of categorizing others into ingroups and outgroups is enough to foster intergroup bias and, conversely, to reduce it when group boundaries are rearranged or questioned.” The 59 examples of social categorization they identified had an average effect size of 0.37; those in the highest quintile of participant size (and thus greatest statistical power) had an average effect size of 0.31.

Examples of activities that investigate who belongs in physics include Careers in Physics, from the STEP-UP curriculum, which emphasizes that physics careers can be focused on solving societal problems, and encourages students to envision themselves in a physics career. The case studies of professional physicists are particularly powerful tools to help participants expand their understanding of who belongs in physics. Another example from a physics setting is to Google “famous physicists” and then critically explore the results (which are typically less diverse than even the very low levels of diversity in physics) [38].

My take: If your primary concern in carrying out a DEI intervention is to improve the experiences of department members from minoritized groups, then asking participants to consciously explore who belongs in physics is an important lever in trying to bring more kinds of people in.

### E. Facilitate discussion and reflection

*Group discussions can reveal group norms and create the opportunity for peers to influence one another* [35]. *One way to make sure the norms and influences are pro-DEI is to empower group members with strong DEI commitments to lead the discussion; another is to use documents like the 2016 Open Letter to SCOTUS from Professional Physicists to set the tone that physicists value inclusion* [38,55].

Group discussion has been shown experimentally to be a modestly effective DEI intervention tactic. Paluck and colleagues found that interventions incorporating discussion and reflection had effect sizes of 0.2–0.27 [30].

This activity was one of only two that Hazari and colleagues found to be associated with increasing women's physics identity [63]. The reasoning behind the effectiveness of this activity is that it reveals group norms and allows for the exercise of peer influence. Gill and Olsen point out that being exposed to anti-bias norms "broadly, from one's own group, or from leaders, can serve as external motivations to reduce prejudice and get back in step with the group. People are particularly likely to adjust their behaviors to the group norms when they identify strongly with the group" [35].

The effectiveness of group discussion is likely to increase when group members who are already committed to DEI are leading discussions, particularly when these group members are socially prominent. Another activity that can work in physics settings is to assign the 2016 Open Letter to SCOTUS from Professional Physicists to students before discussion. This document conveys the anti-prejudice norms held by many physicists and helps forestall the "but why are we doing this in physics?" argument by establishing the value that at least 2463 physicists place on inclusion [38,55,100].

My take: Discussion not only helps participants explore their own ideas, but they may also learn that commitments to bias reduction are more widely shared in their environment than they had realized.

### F. Invoke values, beliefs, and self-affirmation

*Opening DEI interventions with self-affirmations, reminders of individuals' and groups' values, and invitations to participants to reflect on their own values can make the interventions more successful* [30,35,48]. *This works by reducing participants' defensiveness and by reminding them that they can work together to solve a problem that violates their values* [112].

Instead of opening a DEI intervention with statistics, "consider first engaging people in an exercise allowing them to reflect and affirm themselves, or highlighting positives about the organization and its employees that provide this sense of affirmation" [48]. Research indicates that this works by reducing defensiveness [112] and "reminding individuals of their own or their group's values or past egalitarian behaviors, leveraging preferences for self- and group-based consistency to urge individuals toward an anti-prejudice stance" [30]. Gill and Olson cite research findings that after self-affirmation exercises, white people "more readily acknowledged racism" and the role of white people in perpetuating it [35]. Paluck and colleagues found effect sizes of 0.29–0.41 for activities that incorporated values consistency and self-worth [30].

This can consist of "self-affirmation, in which people are invited to reflect on a personally important trait, value, or achievement, why it is important to them, and how it is expressed in their life" [48]. This might look like reminders of "individuals' or their group's egalitarian preferences or history in order to inspire consistency with that history in the present moment" and "moral exemplars" (like the people who co-signed the 2016 letter); encouragement to reflect on "one's existing beliefs and prejudices" and "the prejudice confrontation intervention, in which individuals are given feedback about their level of prejudice" [30]. I have done something like this myself by asking participants at DEI interventions to answer "What do you like about physics?" and then "What do you like about the physics community?" to an anonymous online platform (I used Ref. [113]) and then up-voting answers they agreed with. This created a sense among participants of a shared identity; it reminded participants that they have much in common, which seemed to make it easier to consider how their experiences differed by race and gender, and to take action to create a more inclusive community [56]. I would add that the answers were delightful, and also underscored my own affection for physics and physicists.

### G. Create opportunities for action

*This last one is not an activity. It's an action that I advise leaders in physics departments and workplaces to take if they want their DEI interventions to be successful: Addressing cultural norms of the workplace that interfere with anti-bias actions. Norms that can lead to stress, time pressure, poor sleep, burnout, and substance abuse limit group members' opportunities to take the actions that are the goals of most DEI interventions* [35]. *Leaders who are committed to creating a more inclusive institutional culture would be wise to institute cultural norms that decrease members' stress.*

This last activity is less something to be included in a DEI intervention and more something that leaders can institute in order to make interventions more successful: Create more opportunities for actions that overcome bias. Gill and Olson call this the "opportunity factor" and report that "[e]xisting analyses of DT devote attention to bias and motivation, but components of DT that would fit within the opportunity factor have received little attention" [35]. They identify five factors that interfere: stress, time pressure, poor sleep, burnout, and substance abuse. They note dryly that "U.S. workers are likely to be in a chronic state of reduced opportunity."

Although creating action opportunities may seem daunting, I call on my colleagues in physics to consider ways they could reduce the stress and time pressure that marks their institutions, as those two levers can, in turn, increase the amount of sleep that members of their institutions get, reduce the likelihood of burnout, and perhaps even decrease the likelihood of substance abuse. Physics classes, for example, are often made unnecessarily hard, particularly for members of minoritized groups [114]. Physics problem sets are hard enough; having to cope with unsupportive faculty members and a "culture of no culture" (which is actually a culture characterized by the belief that physics is a true meritocracy, where the best will succeed with no regard to gender, race, or any other personal characteristics [115]) adds unnecessary stress. As Ong puts it, "[a]lthough these norms can be opaque and obscured under ideas about scientific objectivity and neutrality, they become highly visible when challenged, questioned, or broken" [71]. In her book *The Double Bind in Physics Education: Intersectionality, Equity, and Belonging for Women of Color* [71], Ong lays out a series of steps that university physics departments and physics workplaces can take to create a more inclusive culture; ironically, these steps will also increase the opportunity for members of those settings behave more inclusively themselves, by reducing their stress and time pressure, and thus potentially their burnout, poor sleep and substance abuse:

1. "Acknowledge that physics has a culture; allow its norms to be questioned and changed" by conducting self-studies or external reviews, and acting on their recommendations.
2. "Enact accountability for all forms of identity-based harassment and stereotyping," which can dramatically lower the stress of people from groups that are underrepresented in physics.
3. "Listen to women of color and other marginalized students" to learn what is already working in one's setting and what could be changed.
4. Provide funding for women of color and members of other underrepresented groups to participate in identity-based STEM organizations; again, this can lower the stress experienced by some members of physics communities.
5. "Make room for multiple cultures and whole selves inside physics spaces" by acknowledging and even valuing cultural differences within physics.
6. Broaden what it means to be successful in physics. Ong's participants "learned from their professors that pursuing academic physics research was considered to be the singular path to 'success' in their field." This path involved "cutthroat competition, lack of recognition of their intellectual contributions, expectations to work all the time, realizations that this role meant less money and more stress relative to other options, and little recourse to addressing nearly daily discrimination and harassment." Changing this cultural norm in a physics setting to one in which being good in physics includes being a successful teacher, finding a well-paying job in industry, having room for friendship, family, and hobbies, and infusing one's actions in physics settings with one's values would create enormous opportunities for physicists to take antibiased action.

## XIII. HOW WILL YOU KNOW WHETHER YOUR DEI INTERVENTION WORKED?

In this section, I present some questions that will help you systematically evaluate your DEI intervention; see Table IV for specific evaluation items other researchers have used to measure various outcomes. Most DEI programs are not systematically evaluated. Based on their review of around 250 DEI interventions, Devine and Ash report that research reports on the effectiveness of DEI interventions "too often use proxy measures for success that are far removed from the types of consequential outcomes that reflect the purported goals of such trainings"; thus, these reports do not "allow us to reach decisive conclusions regarding best practices in diversity training" [28]. They argue, and I agree, that we are ethically obliged to demonstrate that, at minimum, DEI interventions do no harm and, ideally, meet their intended goals. Otherwise, we are not only wasting participants' time but possibly making situations worse for the very people the interventions were designed to support. I am passionately committed to DEI work, but I deeply resent being forced to attend poorly designed interventions; I can only imagine the kind of resentment that reluctant participants would feel. It is not hard to imagine that resentment being turned not toward the people who required the intervention but toward minoritized members of physics settings.

### A. What should you measure?

*I recommend against using changes in implicit bias as your sole outcome measure. Instead, I recommend that you identify the most important outcomes you would like to achieve with your intervention (this is the subject of the section above called "Why are you doing this intervention? What outcomes do you hope for?") and choose measures that match your intended outcomes. This might mean measuring participants' changes in awareness of bias and its impacts, use of strategies to overcome bias, opportunities to use those strategies, willingness and motivation to reduce bias, increases in participants' physics identity or interest in pursuing a physics career, and an increase in participants' sense of belonging in physics settings.*

Despite the ease in measuring changes in implicit bias, I would recommend against using this unless it is just one of several measures, given the finding that changes in implicit biases do not correlate with changes in explicit biases or with intention to behave differently. Instead, good outcomes to measure could include changes in participants' awareness of diversity issues, changes in their explicit biases, and changes in their "propensity to take action on diversity issues" [22]. Another excellent long-term measure would be whether participation in the DEI intervention changed "the hiring, retention, and perceived belonging of employees from historically marginalized groups over time" [28]. In longitudinal studies, researchers were able to show that a habit-breaking intervention resulted in college students' willingness to take anti-bias actions two years after the intervention [59], and a gender-bias habit-breaking intervention led to the hiring of more women faculty two years later [62]. These results are impressive and make a compelling argument about why it is worth participating in similar interventions.

For those who are interested not just in whether their DEI intervention achieved its goals but in the study of interventions themselves, another area to examine would be "the intervention's efficacy with different groups" [22]. Several of the DEI interventions I collected in the Appendix took this approach; researchers were able to verify that their interventions were not harmful to minoritized participants [42,55,56,64]. One intervention turned out to have a bigger impact on women students than others [110]. These intersectional disaggregations of data provide powerful information to help design, choose, and fine-tune interventions that will better meet our institutional goals.

In deciding what data to gather to determine how well your intervention worked, you should consider both quantitative and qualitative data. Because one of my primary concerns in writing this paper was to determine the magnitude of the effects of DEI interventions, I have focused on quantitative studies to make many of my arguments. That said, I am a qualitative researcher and find the nuance available in qualitative data to be powerful and convincing.

In Table IV, I have assembled quantitative instruments that researchers have used to measure a variety of outcomes; my goal with this table is to make it easy to find at least one way to measure each of the outcomes I discussed in the section above, "Why are you doing this intervention? What outcomes do you hope for?" For the richest data to help you understand whether your intervention met your goals, I suggest you consider pairing quantitative measures with carefully designed open-ended questions or written statements about participants' views on the intervention.

### B. How should you measure outcomes?

If the outcome variables you have chosen can be measured quantitatively, two straightforward ways to determine whether your intervention worked are (i) a pre-post design; (ii) an experimental (or quasiexperimental) design. In a pre-post design, you administer the same survey to participants before and after the intervention, and investigate whether their responses change. (You can also administer the survey a few weeks or months later, to see whether changes were short-term or more long-lasting.) In a quasi-experimental design, you administer the survey to two similar groups of people, one that participated in the intervention and another group that did not. For example, you might administer the survey to two introductory physics classes, only one of which participated in the intervention. (In an experimental design, participants are randomly assigned to the participant and control group; since students weren't randomly assigned to physics classes, this is a quasiexperiment).

Whether you use a pre-post design or a quasi-experimental or experimental design, you will want to investigate two questions: (i) Is the difference between the two sets of scores big enough that it's not likely to be due to chance? (ii) Is the difference big enough to be worth the effort of carrying out the intervention?

We answer "is this difference just due to chance?" by measuring the $p$ value—the probability of obtaining the observed results even if there were no true difference between the two groups. Typically, social scientists accept a $p$ value of less than or equal to 0.05 (5% probability that the difference was due to chance) as acceptable ("statistically significant"). With a simple design, where you are just comparing the scores on a preassessment to those on a postassessment, or the scores on the same assessment taken by two different groups, the simplest statistical test to use is the $t$ test, which calculates the $p$ value using the means, standard deviation (a measure of the variability in the data) and sample size of two sets of scores. The interpretation of $p$ values is tricky, as they are strongly impacted by sample size. As the authors of the American Sociological Association statement on $p$ values put it, "Any effect, no matter how tiny, can produce a small $p$ value if the sample size or measurement precision is high enough, and large effects may produce unimpressive $p$ values if the sample size is small or measurements are imprecise" [116]. In the context of assessing whether a DEI intervention in physics worked, we are, unfortunately, unlikely to face the problem of too many participants. Our problem is more likely to be too few participants to detect real differences. A rule of thumb is that to avoid missing real effects, we need a sample size of at least 30 in each group—either at least 30 participants with both preintervention and postintervention scores, or at least 30 participants in the control group and in the intervention group [117]. Further, if your intent is to determine whether your intervention was worth doing, you might consider using the $p$ value to give you a sense of your data, rather than using a strict 0.05 cut-off. So a $p$ value that is in the vicinity of (but not less than) 0.05, especially with a small sample size, could indicate that your intervention is worth further exploration.

As to the second question—whether the effect is big enough to be worth the effort—this is determined by measuring effect size. Effect size, happily, is independent of sample size, at least when measured using Cohen's $d$: $d = (M_1 - M_2)/(\text{pooled standard deviation})$, where $M_1$ and $M_2$ are the means of the experimental and control groups or prescores and postscores. (There are other common measures of effect size that are more appropriate with other measures [118]). A standard rule is that an effect size of 0.2 is small, 0.5 is medium, and 0.8 is large [34]. However, as I discussed in Sec. V, DEI interventions typically yield small effect sizes; under the right circumstances, these modest effects can lead to bigger impacts.

If you are collecting qualitative data (and, as a qualitative researcher myself, I hope you are), your goal is to look for patterns in the data. Qualitative data let you answer more subtle questions, including "what did my participants experience during this intervention? Was it what I hoped they would experience?" and (assuming your qualitative data suggest that the intervention was effective) "why did this intervention work?" Qualitative findings can also bring outcomes to your attention that you did not even think to measure; participants may describe a particular experience or mention an insight you did not think to ask about in your quantitative instrument, which proves to be important in understanding how your intervention worked or improving it. Qualitative findings can also help you determine whether the intervention had different impacts on different kinds of students. In an example that illustrates several of these points, my undergraduate research assistant was interviewing juniors and seniors about their experiences majoring in physics, and without even asking about a DEI intervention they had participated in several years prior, more than half of her participants brought up the intervention. All the women and some of the men in her sample described it as benefiting them personally, and most of the remaining men described it as valuable even if they didn't feel they had benefited [56].

The tricks to good qualitative research are (i) good questions; (ii) rigorous analysis. Despite my experience with this (or perhaps because of it), I find myself at a loss as to how to write a few paragraphs on how to do these things. Instead, I direct you to the excellent 2009 and 2023 articles by Otero *et al.* (for the 2023 chapter) on doing qualitative physics education research [119,120]. To understand how qualitative physics education researchers make sense of their findings, I point you to the article by Robertson *et al.*, in which they synthesized interviews with qualitative researchers and literature on qualitative research to lay out the logic on which claims about qualitative research rest [121].

My take: In assessing your DEI intervention quantitatively, I encourage you to measure both significance and effect size, but put the significance into context by considering the sample size, and put the effect size into context by considering the typically small effect sizes of similar interventions. Consider including a qualitative component as well, as the results can have powerful explanatory power.

### C. What quality of evidence will you use?

*If you just want to know whether your DEI intervention met your intended outcomes, you can gather data matched with the outcomes and use your findings to further develop your intervention. If you would like to publish your findings, you could consider using an experimental and a control group. If you would like to pursue the highest level of rigor in evaluating your intervention, a control group is required and I suggest you consider using preregistraton (where you submit your research questions and data gathering and analysis plans to a repository ahead of time) and even registration (where your research questions and plans are peer-reviewed before you carry out the research, with the promise of publication even if your study indicates that your intervention was not successful).*

To determine the quality of evidence you will gather, consider the purpose of your evaluation. Is your goal to:

- Determine for yourself and others whether your intervention met its intended goals?
- Pursue publication of your findings?
- Achieve the highest level of rigor?

If your goal is to determine whether your intervention met its intended goals, you will want to match the data you gather to your stated outcomes, although this may require changing your outcomes or gathering complex data. For help with measuring particular outcomes, see Table IV, where I have listed possible outcomes and corresponding outcome measures.

If your goal is to publish findings on the efficacy of your DEI intervention, you may want to consider using an experimental design, in which people are randomly assigned to experience or not experience the intervention. This is not necessary for publication; many of the DEI physics interventions I gathered in the Appendix as exemplars did not use an experimental design. However, use of an experimental group and a control group will result in more convincing, more persuasive findings; if you have a DEI intervention that you have found through evaluation to be promising, testing it out using an experimental design will give you powerful insights into whether it is replicable.

Finally, if you want to achieve the highest level of rigor in evaluating your intervention, this will require an experimental design, and you would be wise to preregister your design and register your study [122]. Preregistration is the process in which researchers "define the research questions and analysis plan before observing the research outcomes" and submit their plans before gathering data. The goal of preregistration is to develop a hypothesis *a priori*, to protect yourself against the temptation to gather data, draw conclusions that fit the data you gathered, and then present the study as though you had originally hypothesized the outcomes you actually found. While this seems ludicrous when you read about it (the basis of scientific thinking is to generate and then test a hypothesis, right?), in practice, it is much murkier, particularly with data gathered about human beings, which is notoriously complicated. The cycle of preregistration goes like this: Consider existing data; develop a hypothesis and a plan to gather and analyze data to test that hypothesis; register the hypothesis and plan; then carry out the study. This process of preregistration offers credibility to your study design and findings. Preregistrations can be submitted to a preregistration repository created by the Center for Open Science [123].

The process of registering your study goes a step beyond preregistration; it allows for peer review of your data-gathering methods before you collect and analyze the data.

TABLE IV. Quantitative tools to measure intended outcomes. Other good sources for instruments and scales can be found in the test collection at ETS [124] and APA PsycTests [125].

| Intended outcome | How others have measured this outcome |
| --- | --- |
| Reduce implicit bias | Pre-post scores on the Implicit Association Test [126]; see also Project Implicit |
| Increase participants' awareness of their own or others' bias | Modern sexism scale [127,128], which includes items like "Discrimination against women is no longer a problem in the United States" and "Over the past few years, the government and news media have been showing more concern about the treatment of women than is warranted by women's actual experiences"<br>"I notice when others exhibit bias towards …"<br>"I could unintentionally behave in biased ways towards …"<br>• "Individuals from racial/ethnic minority groups"<br>• "Women"<br>• "Individuals from any minority group" [129] |
| Increase participants' awareness of the impacts of bias on others | "I consider discrimination against … to be a serious social problem"<br>"… are overly sensitive about unintended offenses"<br>• "Individuals from racial/ethnic minority groups"<br>• "Women"<br>• "Individuals from any minority group" [129] |
| Teach participants strategies to overcome bias | "I am confident I can recognize when bias is occurring during an interpersonal interaction"<br>"I am confident I can speak about equity and diversity in my workplace to my colleagues"<br>"I am confident I can challenge a personnel decision if I think it has been influenced by stereotypes"<br>"I am confident I can challenge a clinical decision if I think it has been influenced by stereotypes"<br>"I am confident I can intervene if I witness a student, resident, fellow, or colleague being treated in a biased way"<br>"I am confident I can assess my office décor, clinic décor, division website, and/or teaching materials for language or images the reinforce negative stereotypes"<br>"I am confident I can adopt the perspective of a student/resident/colleague who is a minority group member"<br>"I am confident I can become better acquainted with a person whose background is different from my own" [129] |
| Increase opportunities to practice strategies to overcome bias | Investigate whether, after the intervention, there is evidence that the institution has committed to "reduce excess duties/obligations of those making hiring, evaluation, and promotion decisions" [35]<br>Longitudinally, investigate whether patterns (perhaps hiring patterns, graduate student admissions, how funding is allocated) have become more equitable [31]<br>"I recognize when bias is occurring during an interpersonal interaction on a regular basis"<br>"I speak about equity and diversity in my workplace to my colleagues on a regular basis"<br>"I challenge a personnel decision if I think it has been influenced by stereotypes on a regular basis"<br>"I challenge a clinical decision if I think it has been influenced by stereotypes on a regular basis"<br>"I intervene if I witness a student, resident, fellow, or colleague being treated in a biased way on a regular basis"<br>"I assess my office décor, clinic décor, division website, and/or teaching materials for language or images the reinforce negative stereotypes on a regular basis"<br>"I adopt the perspective of a student/resident/colleague who is a minority group member on a regular basis"<br>"I become better acquainted with a person whose background is different from my own on a regular basis" [129] |
| Increase motivation to reduce bias/ willingness to take action | Promotion vs prevention scale. Promotion items included:<br>"Right at this minute, in terms of my approach to diversity, I'm feeling … free to pursue my goals/confident that I can go after my goals/focused on what I will achieve"<br>Prevention items included:<br>"Right at this minute, in terms of my approach to diversity, I'm feeling … as though I need to avoid risks/like I don't want to make any mistakes/like I want to make sure nothing bad happens" [45]<br>General motivation items:<br>"I want to recognize when bias is occurring during an interpersonal interaction"<br>"I want to speak about equity and diversity in my workplace to my colleagues"<br>"I want to challenge a personnel decision if I think it has been influenced by stereotypes"<br>"I want to challenge a clinical decision if I think it has been influenced by stereotypes"<br>"I want to intervene if I witness a student, resident, fellow, or colleague being treated in a biased way" |

*(Table continued)*

TABLE IV. *(Continued)*

| Intended outcome | How others have measured this outcome |
| --- | --- |
| | "I want to assess my office décor, clinic décor, division website, and/or teaching materials for language or images the reinforce negative stereotypes"<br>"I want to adopt the perspective of a student/resident/colleague who is a minority group member"<br>"I want to become better acquainted with a person whose background is different from my own"<br>Internal motivation item: "When I promote equity in my division, I do so because of my personal values"<br>External motivation item: "I only go along with my division's diversity goals because everybody else is" [129] |
| | Willingness to take action:<br>Ask participants to envision and write down what actions they will take as a result of the intervention [25,31] |
| | Taking action:<br>Months or years after participating in the intervention, present participants with a vignette that would allow them to take anti-bias actions (or not) and collect their responses; for example, Forscher and colleagues presented participants in their experimental and control groups with a hypothetical essay advocating for stereotyping, and compared the numbers in each group who posted comments disagreeing with the essay [59] |
| Increase physics identity/intent to pursue a physics major or career | Physics identity: "Do you see yourself as a physics person?" [130]<br>Intent to pursue a career in physics: Ask students to "rate the likelihood of their choosing a career in the physical sciences from '0—Not at all likely' to '5—Extremely likely'" [63] |
| Increase belonging/decrease isolation of physicists who are members of minoritized groups | "I feel like I belong in this class"; "I feel like an outsider in this class"; "I feel comfortable in this class"; "Sometimes I worry that I do not belong in this physics class" [73]<br>"With respect to a physics community, to what extent do you … feel different from others in the community?" "With respect to a physics community, to what extent do you … feel alone or isolated?" "With respect to a physics community, to what extent do you … feel inadequate as a member?" [131] |

A limited (but growing) list of journals offers the option to register your study before carrying out the research. When a journal accepts an article after the registration review stage, the understanding is that they will accept the report on the completed study, barring "failings of quality assurance, following through on the registered protocol, or unresolvable problems in reporting clarity or style" [123]. This means the journal will publish the article reporting on the study, regardless of whether the study finds what the researchers' expectations or disconfirms the hypotheses or research questions. The goal here is not only to improve the quality and transparency of research methods but also to avoid publication bias. Publication bias is the tendency to publish only studies that show significant findings, thus preventing us from knowing how many studies may have shown that something did not work. Those of us who want to design effective DEI interventions have a vested interest in learning about activities that did not work. Yet publication bias often prevents us from knowing about how often a category of activity fails, as most published research focuses only on instances where the intervention worked. Designing and carrying out a registered study protects researchers by ensuring that, as long as they conduct their research with fidelity and report it clearly, their efforts will receive recognition through publication. Registration also protects the rest of the broader community by ensuring that we will learn both about instances where an intervention worked to promote DEI in a setting and instances where it did not work. Although preregistration and registration require time and commitment from researchers, they yield the highest-quality evaluations.

## XIV. IN CONCLUSION

Carter, Onyeador, and Lewis wrote in 2020 that "diversity training should go beyond telling people that bias exists or creating uncomfortable experiences that are more likely to prompt defensiveness than learning. Rather, the most effective training is anti-bias training that is designed to increase awareness of bias and its lasting impact, plant seeds that inspire sustained learning, and teach skills that enable attendees to manage their biases and change their behavior." I hope that the information I have gathered in this paper will help physicists who want to push against this tide and create environments that are more diverse, inclusive, and equitable; environments where all of us can thrive.

## DATA AVAILABILITY

No data were created or analyzed in this study.

## APPENDIX: EXEMPLARS

In this appendix, I chose 11 physics and STEM interventions to analyze using the seven questions I explored in this article. Exemplars were chosen either because they were carried out in physics settings with evidence that was at least sufficient for publication, or, for exemplars carried out in other STEM settings, because they used an approach I thought would transfer well to a physics setting. Although many of the exemplars are evaluated based on participant self-report, the researchers evaluated the interventions on the intended outcomes (rather than, for example, on participant enjoyment of the intervention, which Devine and Ash reported was common among the 250 DEI interventions they studied) [28]. I located only two reports that used the highest level of rigor (an experimental design with a control group). I did not locate any research on the effectiveness of DEI interventions that used preregistration or registration. I strongly urge researchers to consider experimental designs and preregistration or registration of their research to help guide DEI efforts in physics. In the table below, I answer the following questions about each exemplar:

1. Who were the participants? How did the researchers take resistant participants into account? What about participants from minoritized groups?
2. What was the intended outcome of the intervention?
3. What form did the intervention take? A workshop? A unit that takes several days within a physics course? A theme that runs throughout a physics course? (for definitions of each, see Table II).
4. How did researchers justify the intervention to participants? (In some cases, this information was missing from published reports, so it is not included in the tables below.)
5. Was the intervention voluntary or required?
6. What activities did it consist of?
7. Why did researchers think the activities would lead to their intended outcomes?
8. How was the outcome measured?
9. What was the quality of their evidence?

| Workshop: Scientific diversity intervention [45] | |
| --- | --- |
| Who were the participants? | Faculty who were already planning to attend a National Academies Summer Institute<br>Resister considerations: Participation was voluntary; the workshop emphasized shared responsibility for inclusion rather than assigning blame |
| Intended outcome | Increase participants' awareness of bias and the impacts of bias; teach participants strategies to overcome bias |
| Form? | Two-hour workshop during a summer professional development conference |
| Voluntary or required? | Voluntary |
| Activity | Interactive presentations about research on implicit biases and the benefits of heterogeneous learning environments, followed by small and then large group discussions, after which participants practiced evidence-based strategies, including mentoring approaches, difficult conversations, and varied teaching approaches |
| Why did the researchers think the chosen activities would lead to the intended outcome? | Emphasizing shared responsibility (rather than individual blame) for equity → participants taking action (rather than avoiding action in order to avoid blame) |
| How was the outcome measured? | Effectiveness was measured through a pre-post self-report survey. Participants were more aware of gender bias, expressed less gender bias, and were more willing to engage in actions to reduce gender bias 2 weeks after participating in the intervention, compared with 2 weeks before the intervention |
| What was the quality of evidence? | Sufficient for publication: Participant self-report evaluating intended outcomes |
| **Workshop: Breaking the gender bias habit [31]** | |
| Who were the participants? | Faculty in medicine, science, and engineering departments at UW-Madison<br>Resister considerations: Participation was voluntary, and the workshop focused on how everyone could break gender bias habits rather than suggesting that one group was responsible |
| Intended outcome | Teach participants strategies to overcome bias: Replace gender bias habits with gender equity habits |
| Form | The 2.5-h workshop was intended to increase participants' awareness of bias and the impacts of bias; teach participants strategies to overcome bias: increase motivation and willingness to take action |
| Justification? | The "business case"—that better science is done by a more diverse group of scientists |
| Voluntary or required? | Voluntary |

*(Table continued)*

*(Continued)*

| | |
|---|---|
| Activity | Presented research on bias as a habit and on "bias literacy" (how group stereotypes affect how individuals are perceived; how cultural assumptions around gender can lead to "social penalties" for violating those assumptions; how men can benefit from being in settings associated with male stereotypes; and stereotype threat). Presented behavioral strategies that can help overcome bias (stereotype replacement, positive counter stereotyping, perspective taking, evaluating people as individuals, and increasing contact with a wide range of people) and strategies that don't work (gender blindness; relying on one's own purported objectivity). This was followed by small and whole group discussions, case studies, and "a written commitment to action" |
| Why did the researchers think the chosen activities would lead to the intended outcome? | Increasing faculty awareness of the impacts of gender bias and of effective (and ineffective) strategies for overcoming bias → increased motivation and self-efficacy → automatic action without bias |
| How was the outcome measured? | This study included 92 departments, randomly assigned to treatment or control groups. After the workshop, faculty (both male and female) in the treatment departments differed significantly from those in the control departments in the degree to which they felt they fit into their department, felt that their colleagues valued their research and scholarship more, and felt more comfortable bringing up family and personal considerations when scheduling work responsibilities. In a follow-up preregistered study, the proportion of women hired by control departments did not change, whereas it increased by 18% in the treatment departments |
| What was the quality of evidence? | Strongest. Preregistered, randomized, controlled study evaluated on intended outcomes |
| Workshop: Making physics stronger [56] | |
| Who were the participants? | Students in introductory university physics classes for physics majors<br>Underrepresented considerations: The workshop was designed to build on participants' shared identity as members of the physics community |
| Intended outcome | Increase motivation to reduce bias and willingness to take action; increase belonging and decrease isolation of physicists who are members of minoritized groups; acclimate new physics majors to an inclusive departmental culture |
| Form | 1–1.5 h workshop |
| Justification? | The fairness case: Physics is an inherently interesting field, but some people face obstacles due to factors outside their control and not linked to their aptitude or willingness to work hard |
| Voluntary or required? | Required; activities occur during class sessions |
| Activity | Teach about inequity; create real or imagined contact; teach strategies to overcome bias; investigate who belongs; facilitate discussion and reflection; invoke values, beliefs, and self-affirmation<br>One-hour workshop including information on: beliefs about "natural ability" and the percent of women and Black people holding Ph.D.s in those fields [91], implicit bias and how its effects can be seen in decision making by scientists [132], stereotype threat. Small group discussions of five case studies from women facing bias in physics settings. Participants were asked to come up with actions they might take if this situation happened to them, to someone they knew, or in front of them while they were bystanders |
| Why did the researchers think the chosen activities would lead to the intended outcome? | Emphasizing the shared physics identity of workshop participants and exploring narratives about the discrimination experienced by some physicists → increased motivation to make physics environments more inclusive |
| How was the outcome measured? | In total, 26 workshop participants were interviewed for another project on student-student interactions in physics. Of these, 16 of them spontaneously brought up their experiences with the workshop. Nine students expressed strong, specific appreciation, including all the women, trans folks, and nonbinary folks, as well as several men. Five more men expressed approval for the workshop in ways that were more moderate and vague. Finally, two men expressed intense, specific dislike of the workshop |
| What was the quality of evidence? | Sufficient for publication: Longitudinal self-reports on perceived value of activity |

*(Table continued)*

*(Continued)*

| | |
|---|---|
| Unit: Underrepresentation curriculum [39,43,47,133] | |
| Who were the participants? | Students in high school and university STEM classes (with a focus on physics)<br>Resister considerations: Instructors are encouraged to pay attention to their own skepticism and that of their students as a way to find effective ways to address it<br>Underrepresented considerations: “When in doubt, we encourage instructors to consider prioritizing the needs of the marginalized over the needs of the majority” [47] |
| Intended outcome | Increase participants’ awareness of bias and the impact of bias; increase motivation to reduce bias; increase sense of belonging of minoritized groups. Change the culture of STEM by teaching about injustice |
| Justification? | Justified to students by telling them that learning about the culture of physics is a part of learning physics |
| Voluntary or required? | Required; activities occur during class sessions |
| Form | Unit within high school or university course |
| Activity | Teach about inequity; facilitate discussion and reflection; create opportunities for action<br>The curriculum is composed of three components; time devoted to each is at the discretion of the instructor: (i) the nature of science, demographics of scientists, and why it matters; (ii) implicit bias, affirmative action, meritocracy, and other topics relevant to underrepresentation in science; and (iii) turning knowledge into action (e.g., writing letters to advocate for faculty of color and organizing diverse study groups) |
| Why did the researchers think the chosen activities would lead to the intended outcome? | Using “the tools of science to recognize inequity in science” [133] → taking action to address inequity |
| How was the outcome measured? | Out of 454 students at one high school who participated, only 19 (4%) said that the experience was not worthwhile [41]. Overall, 90% of 155 students at a 2-year college reported that the Underrepresentation Curriculum was a positive experience [39]. |
| What was the quality of evidence? | Sufficient for publication: Self-reports on perceived value of activity |
| Unit: Four-day equity unit [40,55] | |
| Who were the participants? | Students in introductory university physics classes<br>Resister considerations: Instructors intentionally introduced a variety of views on the importance of diversity in physics<br>Underrepresented considerations: Students from underrepresented groups were told ahead of time that the unit was coming up |
| Intended outcome | Shift students’ views on the importance of diversity in physics |
| Form | Unit within university course |
| Justification? | Justified to students using the 2015 “Open Letter to SCOTUS from Professional Physicists,” [100] signed by almost 2500 physicists |
| Voluntary or required? | Required; activities occur during class sessions |
| Activity | Teach about inequity; facilitate discussion and reflection; create opportunities for action<br>Four-day unit that includes the nature of physics; an analysis of how the absence of certain groups of people from physics impacts what gets done in physics; how participating in physics benefits people from groups that are currently underrepresented; how to explain why certain groups are excluded; and what students can do to change this exclusion |
| Why did the researchers think the chosen activities would lead to the intended outcome? | “provide a space for students to explore their own views of what it means to learn and practice physics and how that might be affected by racial inequity” [40] |
| How was the outcome measured? | Overall, 36% of students indicated that as a result of the unit, they had either changed, gained awareness, or solidified their views on equity and inclusion in physics. In total, 90% had positive views of the unit, even if their views did not shift; only 10% were dissatisfied. In addition, seven students of color indicated that the unit had not changed their views but was important for others [55] |

*(Table continued)*

*(Continued)*

| | |
|---|---|
| What was the quality of evidence? | Sufficient for publication. Self-reports on perceived value of activity; evaluating intended outcomes |
| Unit: Two-day unit on representation in STEM [38] | |
| Who were the participants? | Students in a sophomore-level modern physics class (mostly engineering majors)<br>Resister considerations: Used materials from APS, AIP, AAPT, and other sources to make it clear that professional physicists discuss issues of inclusion in physics |
| Intended outcome | Increase physics identity/intent to pursue a physics major. Make classrooms more inclusive and thus diversify the field |
| Form | Unit within university course |
| Voluntary or required? | Required; activities occur during class sessions |
| Activity | Teach about inequity; investigate who belongs and what it means to belong; facilitate reflection<br>Two-day unit: "what is physics?" followed by "who does physics?" The unit included readings, homework, and exam questions. Activities included googling "famous physicists" and reflecting on the results, looking at APS data on the demographics of physics, and reading the "Open Letter to SCOTUS from Professional Physicists" [100] |
| Why did the researchers think the chosen activities would lead to the intended outcome? | Talking explicitly about equity, diversity, and inclusion in physics → increased identity and belonging → academic success |
| How was the outcome measured? | "In the two semesters reported on, 75%–80% of student responses to the unit on surveys were positive, 15%–20% neutral, and 0%–5% negative. The few negative reactions that we have received have primarily included the idea that these topics are important, but they just do not belong in a physics class" [38] |
| What was the quality of evidence? | Sufficient for publication: Self-reports on perceived value of activity |
| Unit: One-day session on underrepresentation of women in physics [109] | |
| Who were the participants? | Students in high school honors physics class |
| Intended outcome | Increase physics identity and intent to pursue a physics major; increase the sense of belonging of members of minoritized groups; increase girls' intention to pursue physics careers |
| Form | Unit within high school physics course |
| Voluntary or required? | Required; activities occur during class sessions |
| Activity | Real or imagined contact; investigate who belongs in physics; facilitate discussion and reflection.<br>Students read a book about early 20th century physics, then wrote essays describing the work of—and challenges faced by—three women physicists, and answering this question: "Is the opportunity to excel in physics any easier now than it was in the early 20th century?" The essay was followed by a class discussion; discussion themes included famous scientists, gendered professions, and classroom experiences. |
| Why did the researchers think the chosen activities would lead to the intended outcome? | Providing girls in physics with opportunities to reflect on and expand their underlying ideas about who does physics and what physics is (their "figured worlds") → greater likelihood to pursue physics further |
| How was the outcome measured? | This article rests on previous research that indicated that discussing the underrepresentation of women strengthens women's physics identity and interest in physics careers; this particular study (i) demonstrated what this can look like and (ii) probed the underlying dynamics that may explain why discussing underrepresentation leads to these outcomes. The researchers found that "the part of the discussion with the greatest apparent impact focused on the experiences of women in both school science and professional science today" |
| What was the quality of evidence? | Sufficient for publication: Case study; self-reports on perceived value of activity |

*(Table continued)*

*(Continued)*

| Unit: Two-lesson unit: STEP-UP [64,108] | |
| --- | --- |
| Who were the participants? | Students in high school physics classes |
| Intended outcome | Increase students' intentions to pursue physics careers, especially for girls and minoritized students; increase the sense of belonging of minoritized groups |
| Form | Unit within high school physics course |
| Justification? | The business case: that physics relies on people of all backgrounds to advance (the authors argue that this is the way to justify the unit to administrators and parents who might raise concerns) |
| Voluntary or required? | Required; activities occur during class sessions |
| Activity | Teach about inequity; real or imagined contact; investigate who belongs and what it means to belong; invoke values, beliefs, and self-affirmation<br>Two lessons, each 1–2 class periods long, both from the STEP UP curriculum: (i) A Careers in Physics lesson that emphasizes communal goals like benefiting society and helping others (including profiles of dozens of people with bachelor's degrees in physics, mostly women and people of color, and how have a broad range of jobs); (ii) A Women in Physics lesson in which students look at statistics about who does physics, generate hypotheses about why underrepresentation persists, and develop goals to support one another |
| Why did the researchers think the chosen activities would lead to the intended outcome? | Using counternarratives to expand students' cultural beliefs about who does physics, and why → more students will study physics |
| How was the outcome measured? | In a study of ten high school teachers in eight states (823 students total), participating in these lessons led to a significant increase in all students' future intentions to pursue physics. Effect sizes were: female-identified students: 0.29; non-female-identified: 0.19; Minoritized racial or ethnic (MRE) group members: 0.24; non-MRE: 0.23. In a second study of 13 teachers in 3 states, randomly assigned to treatment or control groups, again all students in the treatment group had a significant increase in their future intentions (female-identified: 0.29; non-female-identified: 0.28; MRE: 0.30; non-MRE: 0.21). Among those in the control group (who received lessons with comparable pedagogy on the topic "notable physicists"), there was no increase among female-identified students and non-MRE students; non-female-identified experienced a significant increase (effect size of 0.16), as did MRE (effect size of 0.18). |
| What was the quality of evidence? | Strong. Randomized experimental design evaluating intended outcomes. |
| **Semester-long theme: Equity-focused, identity-conscious curriculum [42]** | |
| Who were the participants? | Physics majors in an intermediate-level modern physics class |
| Intended outcome | Increase participants' awareness of bias and the impact of bias; increase the sense of belonging of minoritized groups<br>Disrupt structural barriers and increase feelings of belonging among minoritized students |
| Form | Semester-long theme in university physics course |
| Justification? | "We strive to integrate the equity-focused curriculum with the technical content in order to illustrate that bias and privilege are interwoven into the very fabric of physics, rather than a stand-alone topic" [42] |
| Voluntary or required? | Required; activities occur during class sessions |
| Activity | Teach about inequity; investigate who belongs and what it means to belong in physics; facilitate discussion and reflection; create opportunities for action<br>Weekly assignments and essays on underrepresentation, bias, and privilege in physics. Topics include implicit bias, imposter syndrome, stereotype threat, microaggressions, structural and institutional racism, and white privilege. This culminates in a student poster session on decolonizing physics and astronomy |
| Why did the researchers think the chosen activities would lead to the intended outcome? | Giving students time to reflect on identity and equity topics and create their own hands-on project → increased feelings of belonging among minoritized students; all students will realize that issues of equity are interwoven into physics, not separate from it |

*(Continued)*

| | |
|---|---|
| How was the outcome measured? | Assessed through interviews and pre-post survey data. Of the 11 interviewees, all indicated that they appreciated the curriculum and that it was valuable to white male students, students of color, and white female students. On surveys, a question about the value of a pro-equity curriculum in physics increased significantly between pre- and post-tests |
| What was the quality of evidence? | Sufficient for publication: Pre-post surveys, post-intervention interviews on perceived value of activity |
| Semester-long theme: Including women as characters in lecture examples and problem sets [110] | |
| Who were the participants? | Students in introductory university physics classes for STEM majors |
| Intended outcome | Increase physics identity and intent to pursue a physics career; increase sense of belonging in physics<br>Increase the diversity of physics students and thus diversify the field |
| Form | Semester-long theme in university physics class |
| Voluntary or required? | Required; activities occur during class sessions |
| Activity | Create real or imagined contact; investigate who belongs and what it means to belong in physics<br>The physics problems students completed were set in more diverse settings than typical problems. Students also completed two assignments about women physicists and developed two of their own physics problems, including scenarios that detailed the context and the players in the problems |
| Why did the researchers think the chosen activities would lead to the intended outcome? | Bringing their own backgrounds and interests into physics class → students "connect physics principles with their own sense of belonging and identity" |
| How was the outcome measured? | Assessed with a control group (who were assigned problems out of the textbook and not given the assignments about women physicists) and a treatment group. Both groups designed their own problems. The students—regardless of gender—in the treatment group were 12 times more likely to include women in the problems. That said, women in both sections were more likely to design problems with women characters. Students in both sections indicated that "the DYOP [design your own problem] assignment pushed them to understand physics at a more nuanced level" |
| What was the quality of evidence? | Sufficient for publication: Use of a control group |
| Semester-long theme: Weekly homework assignments and essays [38] | |
| Who were the participants? | Students in a sophomore-level modern physics class<br>Resister considerations: Used materials from APS, AIP, AAPT, and other sources to make it clear that professional physicists discuss issues of inclusion in physics<br>Underrepresented considerations: Students could communicate their responses to readings directly through their instructors, via written reflections, in addition to classroom discussions |
| Intended outcome | Increase physics identity and intent to pursue physics; increase internal or external motivation to reduce bias<br>Make classrooms more inclusive and thus diversify the field |
| Form | Semester-long theme in university physics course |
| Justification? | Justified to students by tying into the larger course theme of "practicing professionalism" |
| Voluntary or required? | Required; activities occur as part of required classwork |
| Why did the researchers think the chosen activities would lead to the intended outcome? | Writing weekly essays → "a more realistic physics identity that they can see themselves in" |
| How was the outcome measured? | The reflection essay has not decreased the percentage of students who choose to major in physics after the course. Many students initially resist the reflection, but according to student comments in the reflections and course evaluations, eventually find it valuable |
| What was the quality of evidence? | Sufficient for publication: Persistence in physics major; self-reports on perceived value of activity |